%% file: ms.tex
\documentclass[iop]{emulateapj}
\begin{document}

\title{THE SLOAN DIGITAL SKY SURVEY STRIPE 82 IMAGING DATA: DEPTH-OPTIMIZED 
CO-ADDS OVER 300 DEG$^2$ IN FIVE FILTERS}

\author{Linhua Jiang\altaffilmark{1,12,13}, Xiaohui Fan\altaffilmark{2,12},
	Fuyan Bian\altaffilmark{3,12,14}, Ian D. McGreer\altaffilmark{2},
	Michael A. Strauss\altaffilmark{4,12}, James Annis\altaffilmark{5},
   Zo\"{e} Buck\altaffilmark{4,6,12}, Richard Green\altaffilmark{2,12},
   Jacqueline A. Hodge\altaffilmark{7,8,12}, Adam D. Myers\altaffilmark{9,12},
	Alireza Rafiee\altaffilmark{10,12}, and Gordon Richards\altaffilmark{11,12}
}

\altaffiltext{1}{School of Earth and Space Exploration, Arizona State
   University, Tempe, AZ 85287-1504, USA}
\altaffiltext{2}{Steward Observatory, University of Arizona,
   933 North Cherry Avenue, Tucson, AZ 85721, USA}
\altaffiltext{3}{Research School of Astronomy and Astrophysics, Australian 
	National University, Weston Creek, ACT 2611, Australia}
\altaffiltext{4}{Department of Astrophysical Sciences, Princeton University,
   Princeton, NJ 08544, USA}
\altaffiltext{5}{Center for Particle Astrophysics, Fermi National Accelerator 
   Laboratory, P.O. Box 500, Batavia, IL 60510, USA}
\altaffiltext{6}{Department of Education, University of California, Santa 
   Cruz 1156 High Street, Santa Cruz, CA 95064, USA}
\altaffiltext{7}{NRAO, 520 Edgemont Road, Charlottesville, VA 22903, USA}
\altaffiltext{8}{University of California, 1 Shields Ave, Davis, CA 95616, USA}
\altaffiltext{9}{Department of Physics and Astronomy, University of Wyoming, 
   Laramie, WY 82071, USA}
\altaffiltext{10}{Department of Physics, Astronomy, and Geosciences, Towson 
	University, Towson, MD 21252, USA}
\altaffiltext{11}{Department of Physics, Drexel University, 3141 Chestnut 
	Street, Philadelphia, PA 19104, USA}
\altaffiltext{12}{Visiting Astronomer, Kitt Peak National Observatory, 
National Optical Astronomy Observatory, which is operated by the Association 
of Universities for Research in Astronomy (AURA) under cooperative agreement 
with the National Science Foundation.}
\altaffiltext{13}{Hubble Fellow}
\altaffiltext{14}{Stromlo Fellow}

\begin{abstract}

We present and release co-added images of the Sloan Digital Sky Survey (SDSS) 
Stripe 82. Stripe 82 covers an area of $\sim$300 deg$^2$ on the Celestial 
Equator, and has been repeatedly scanned 70--90 times in the $ugriz$ bands by 
the SDSS imaging survey. By making use of all available data in the SDSS 
archive, our co-added images are optimized for depth. Input single-epoch frames 
were properly processed and weighted based on seeing, sky transparency, and 
background noise before co-addition. The resultant products are co-added 
science images and their associated weight images that record relative weights 
at individual pixels. The depths of the co-adds, measured as the 5$\sigma$ 
detection limits of the aperture ($3\farcs2$ diameter) magnitudes for point 
sources, are roughly 23.9, 25.1, 24.6, 24.1, and 22.8 AB magnitudes in the 
five bands, respectively. They are 1.9--2.2 mag deeper than the best SDSS 
single-epoch data. The co-added images have good image quality, with an 
average point-spread function FWHM of $\sim$$1\arcsec$ in the $r$, $i$, and 
$z$ bands. We also release object catalogs that were made with SExtractor. 
These co-added products have many potential uses for studies of galaxies, 
quasars, and Galactic structure. 
We further present and release near-IR $J$-band images that cover $\sim$90 
deg$^2$ of Stripe 82. These images were obtained using the NEWFIRM camera on
the NOAO 4-m Mayall telescope, and have a depth of about 20.0--20.5 Vega 
magnitudes (also 5$\sigma$ detection limits for point sources). 
%All the data are released via ({\bf link}).

\end{abstract}

\keywords
{atlases --- catalogs --- surveys}

\section{INTRODUCTION}

Large-area multiwavelength surveys have revolutionized our understanding of
the properties of distant galaxies and quasars, as well as stars in our own 
Galaxy. The Sloan Digital Sky Survey \citep[SDSS;][]{york00} has been a 
pioneer among these surveys in the last decade. The SDSS is an imaging and 
spectroscopic survey of the sky using a dedicated wide-field 2.5 m telescope 
\citep{gunn06} at the Apache Point Observatory. So far the SDSS has obtained 
spectra of more than 1,500,000 galaxies and 160,000 quasars 
\citep{ahn12,ahn13,par12,par13}. While its imaging survey has been completed, 
its spectroscopic survey is still going on in the phase known as SDSS-III 
\citep{eis11}, so these numbers are growing steadily.

The SDSS imaging survey covered a total of 14,555 deg$^2$ of unique sky area
\citep{ahn12}. Imaging was performed in drift-scan mode using a 142 mega-pixel 
camera \citep{gunn98} that gathers data in five broad bands, $ugriz$, on 
moonless photometric \citep{hogg01} nights of good seeing. The effective 
exposure time was 54.1 s. The five broad bands span the range from 3000 to 
11,000 \AA\ \citep{fuk96}. The images were processed using specialized 
software \citep{lup01,sto02}, and were photometrically \citep{tuc06} and 
astrometrically \citep{pier03} calibrated using observations of a set of 
primary standard stars \citep{smi02} on a neighboring 20-inch telescope. 
The photometric calibration is accurate to roughly 2\% rms in the $g$, $r$, 
and $i$ bands, and 3\% in $u$ and $z$, as determined by the constancy of 
stellar population colors \citep{ive04,bla05}. With the so-called 
`ubercalibration' \citep{pad08} which uses overlap between imaging scans, 
the calibration residual errors are reduced to $\sim2$\% in $u$ and 
$\sim1$\% in $griz$.

The majority of the SDSS imaging data are single-epoch images (except
overlap regions between adjacent scans). But in addition to single-epoch 
data, the SDSS also conducted a deep survey by repeatedly imaging a $\sim300$
deg$^2$ area on the Celestial Equator in the south Galactic cap in the Fall
\citep{ade07,ann11}. This deep survey stripe, or Stripe 82, roughly spans 
$\rm 20^h<R.A.<4^h$ and $\rm -1.26\degr<Decl.<1.26\degr$. 
Stripe 82 was scanned 70--90 times, depending on R.A. along the stripe,
in 1998--2007. In 1998--2004, roughly 80 SDSS imaging runs were taken on Stripe 
82, usually under optimal observing conditions mentioned above. In 2005--2007, 
more than 200 additional runs were taken as part of the SDSS-II supernovae 
survey project \citep{fri08}. The observing conditions for many runs in 
2005--2007 were less optimal, with significant moonlight, poor seeing, or 
non-photometric transparancy, as we will discuss in section 2.1.
The multi-epoch images of Stripe 82 are suitable for studies of variability 
and transient events. For example, they have been used for high-redshift 
supernovae survey \citep{fri08,sako08} and quasar variability studies 
\citep[e.g.][]{mac12,sch12}. The multi-epoch data also allow the 
construction of deeper co-added images \citep{aba09,ann11,huff11}.

The first version of co-added images made from the Stripe 82 images are 
publicly available in the SDSS database. They were released in the SDSS Data 
Release 7 \citep{aba09}. The details of the construction of the co-adds are 
described in \citet{ann11}. Briefly, \citet{ann11} combined images taken 
before 2005 December 1, over $\rm -50\degr<R.A.<60\degr$ of the Stripe 82. 
Each area of sky included data from 20--35 runs. The co-added images were then 
run through the SDSS pipeline to generate catalogs and other standard SDSS 
products. The total coverage of the co-added catalog is 275 deg$^2$.
The data are about 1--2 mag (depending on R.A. and bands) deeper than SDSS 
single-epoch data. \citet{huff11} also produced co-added images from the
Stripe 82 data. Their products were mainly used for studies of galaxy weak 
lensing, so they only included images with relatively good seeing. 
In \citet{jiang09}, we made our own co-added images that we used to select 
high-redshift ($z>5$) quasars \citep{jiang08,jiang09,mcg13}.

In this paper, we release a new version of the co-added images and their
associated object catalogs. This version includes all 
available images that cover Stripe 82. The co-add methodology is slightly 
different from that of \citet{ann11}. The main difference is that 
\citet{ann11} released catalogs of detected objects with properties measured 
by the standard SDSS pipeline, but we did not run the SDSS pipeline. Instead, 
we produced object catalogs using {\tt SExtractor} \citep{ber96}. However, our 
images include many more SDSS runs than \citet{ann11}, so they go considerably 
deeper. The additional runs we included were taken in
2006--2007 as part of the SDSS-II supernovae survey, as mentioned earlier.
We also release near-IR $J$-band images that cover about 90 deg$^2$ of 
Stripe 82. These images were obtained from the NOAO Kitt Peak 4-m Mayall 
telescope, using the wide-field 
near-IR imager NEWFIRM \citep{pro04}. They have a depth of roughly 20.0--20.5 
Vega mag (5$\sigma$ detection for point sources), depending on position. This 
is much shallower than the depth of the co-added SDSS images, but represents 
a significant extension of the wavelength range covered by the SDSS filters.

In Section 2, we present the details of the construction of our co-added 
images from SDSS multi-epoch data. We then describe our image products and
quality assessment in Section 3. In Section 4, we present our NEWFIRM $J$-band
data. We summarize the paper in Section 5. Throughout the paper all SDSS
magnitudes are on the AB system (not SDSS asinh magnitudes \citep{lup99};
all asinh magnitudes have been converted to logarithmic AB magnitudes). 
The $J$-band magnitudes are on the Vega system.

\section{CONSTRUCTION OF THE CO-ADDED IMAGES}

In this section, we present the construction of our co-added Stripe 82 images.
While there is no formal definition of the R.A. range for Stripe 82, we adopt 
the range of $\rm -60\degr<R.A.<60\degr$ ($\rm 20^h<R.A.<4^h$) here. This 
range spans Galactic latitudes from $b=-15\degr$ to $b=-63\degr$. The fields 
near $\rm R.A.=-60\degr$ (or $\rm R.A.=300\degr$) are close to the Galactic 
plane, so they are overwhelmed by Galactic stars and dust, and are not 
suitable for extragalactic studies. In addition, these fields have less 
scan coverage (and thus shallower co-added images), as we will see in the next 
section.

\begin{figure} %f1
%\epsscale{1.0}
\plotone{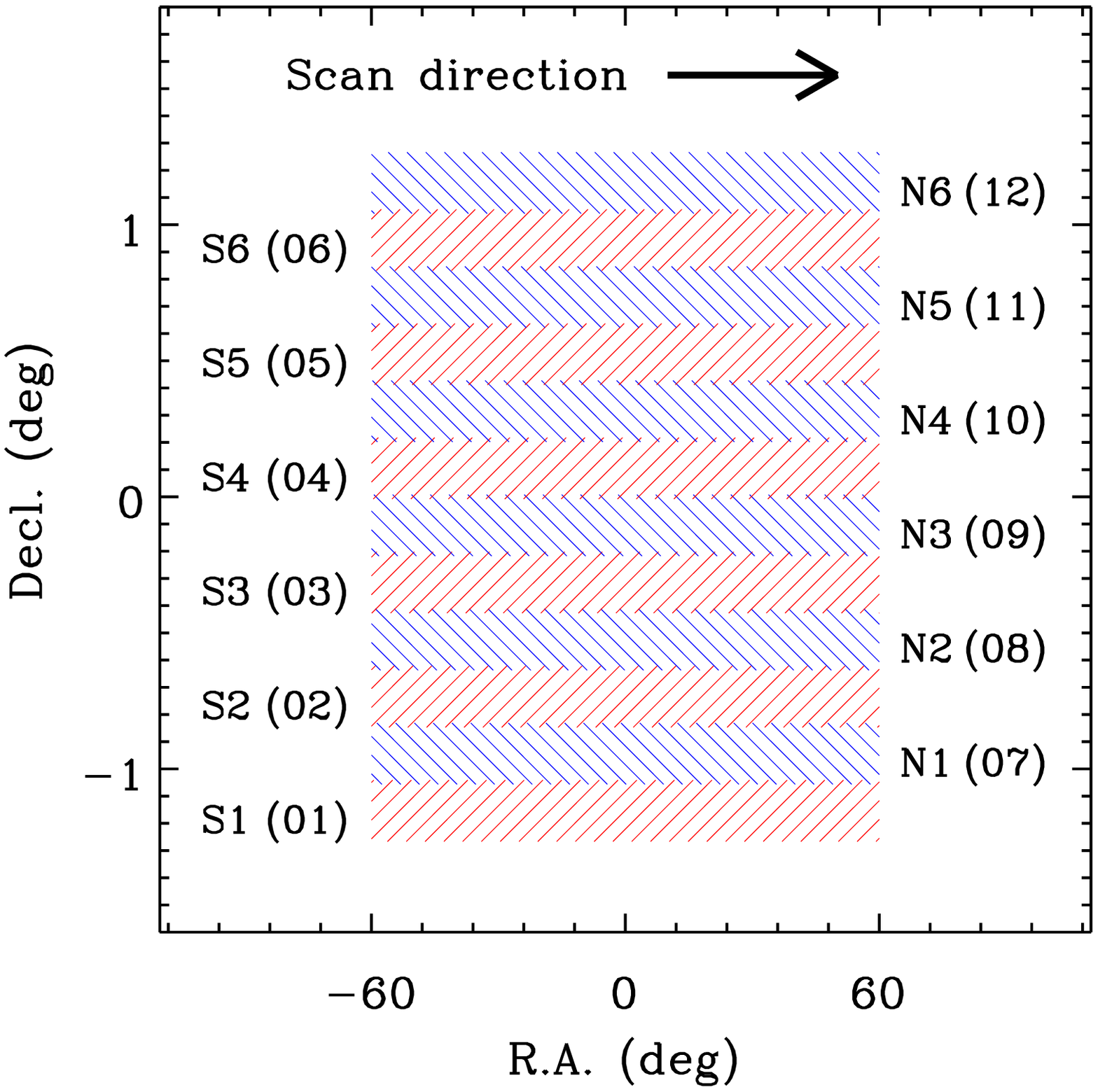}
\caption{Layout of SDSS Stripe 82. Stripe 82 covers
$\rm -60\degr<R.A.<60\degr$ ($\rm 20^h<R.A.<4^h$) and
$\rm -1.26\degr<Decl.<1.26\degr$. It consists of six south (S) scanlines
(red hatch) and six north (N) scanlines (blue backward hatch). The numbers in
the brackets are the scanline numbers used for our co-added images.}
\end{figure}

We started with all 314 runs that cover (part of) Stripe 82. 
An SDSS run (strip) 
consists of 6 parallel scanlines, identified by camera columns or `camcols', 
for each of the five $ugriz$ bands. The scanlines are $13\farcm5$ wide, with
gaps of roughly the same width, so two interleaving strips make a stripe. 
Figure 1 illustrates the two strips (12 scanlines) of Stripe 82, referred to 
as the south (S) and north (N) strips, respectively. In our final co-added 
data, the six S scanlines have scanline numbers from 01 to 06, and the six N 
scanlines have scanline numbers from 07 to 12 (Figure 1). SDSS scanlines are 
divided into fields. An SDSS field is the union of five $ugriz$ frames 
covering the same region of sky, and a SDSS frame is a single image in a 
single band. The size of a field (or a frame) is $1489\times2048$ pixels, or 
roughly $9\farcm8 \times 13\farcm5$ ($\rm R.A. \times Decl.$), with a pixel 
size of $0\farcs396$. There is an overlap region with a width of 128 pixels 
along the scan direction between any two neighbor frames.
The input images for our co-adds are SDSS calibrated frames, or the {\tt fpC} 
images. These frames have been bias subtracted and flat-fielded, with bad and
saturated pixels interpolated over.

\subsection{Run and Field Selection}

\begin{figure} %f2
%\epsscale{0.5}
\plotone{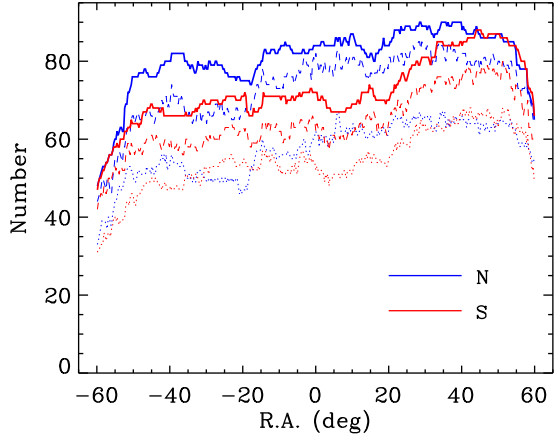}
\caption{Run coverage (number of observing runs/scans) as a function of R.A.
for Stripe 82. The north and south strips are indicated by blue and red,
respectively. The solid lines represent the number of runs chosen for our
co-adds. The median coverage is 82 for the N strip, and 71 for the S strip.
The dashed and dotted lines represent the numbers of runs that were actually
used for the co-adds (scanline 2 in N and S) in the $i$ and $g$ bands,
respectively. See Section 2 for details.}
\end{figure}

As the first step, we chose runs and fields for the Stripe 82 co-adds that
are not of low quality (due to very poor seeing, high sky background, low 
sky transparency, etc.). For each field in the 314 runs between 
$\rm R.A.=-60\degr$ and $60\degr$, we estimated three parameters from its 
$r$-band {\tt fpC} frame: the FWHM of the point-spread function (PSF), the 
atmospheric extinction (or sky transparency), and the sky background. The sky 
background is the median value of the frame, after the artifical soft bias of 
1000 DN is subtracted. This value is consistent with the value provided by the 
keyword `sky' in the fits image header. More sophisticated sky subtraction is 
done later. In order to measure PSF FWHM and extinction, we ran 
{\tt SExtractor} on the {\tt fpC} frame, and measured PSF and flux (in units 
of DN) for isolated bright point sources. We then matched this object catalog 
to the catalog of Stripe 82 standard stars by \citet{ive07}, and computed a 
zero point so that the median magnitude difference between the two catalogs is 
zero. This is the absolute zero point for this frame, regardless of whether
the condition was photometric or non-photometric. So the magnitude of an 
object in this frame is --2.5 log(DN/$\rm t_{exp}$) + zero point, where
$\rm t_{exp}$ is the exposure time of 54 sec. This procedure is the same as 
\citet{ann11} did, except that they used their own standard catalog. 
We found that the median zero point for data taken at photometric nights was 
23.9 mag in the $r$ band, the same as the value given by \citet{ann11}. 
Finally, the relative atmospheric extinction is simply the zero point offset 
from the photometric zero point. For the region that the \citet{ive07} catalog 
does not cover ($\rm R.A.=300\degr - 306.5\degr$), we chose one `good' 
Stripe 82 run with good seeing, low sky background, and nearly zero extinction 
as a standard run. We only took into account the relative extinction for the 
next steps.

\begin{figure} %f3
%\epsscale{0.5}
\plotone{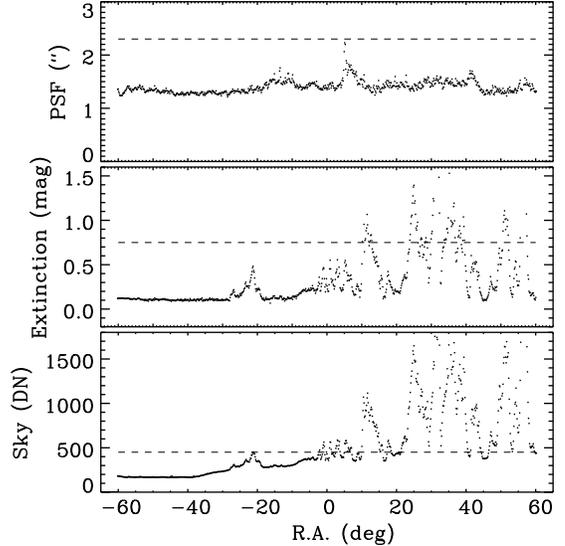}
\caption{Some basic information for Run 7106 in the $r$ band (scanline 6).
From the top to the bottom panels we show PSF, atmospheric extinction, and sky
background for each frame as a function of R.A..
The dashed lines indicate our selection cuts. The image quality at
$\rm R.A.<-40\degr$ in this run is good. The image quality is worse at
$\rm R.A.>-40\degr$, and is unacceptable for the majority of the frames at
$\rm R.A.>0\degr$. Our procedure selected proper frames and weighted
individual frames based on PSF, extinction, and background noise.}
\end{figure}

Based on these three parameters, we rejected 11 runs in which most of the
fields are of very low quality. 
The remaining 303 runs were used for our co-adds. They are listed in 
Table 1 in the Appendix. Figure 2 shows the number of runs as a function of 
R.A. (solid blue and red lines). The median coverages are 82 and 71 for the N 
and S strips, respectively.
We kept runs even if only a small fraction 
of the fields are good. Figure 3 is an example, showing the three parameters
in the $r$ band for Run 7106. The frames at $\rm R.A.<-40\degr$ in this run 
have high quality, with a PSF of $\sim1\farcs3$, low 
extinction of $\sim0.1$ mag, and low sky background of $\sim200$ DN. The image 
quality then gets worse at $\rm R.A.>-40\degr$, and eventually becomes
unacceptable at $\rm R.A.>0\degr$. Our further selection criteria will
select and weight individual frames based on the three parameters.

The SDSS CCDs were read out with two amplifiers. In some runs, the two 
amplifiers had different gains so that the two halves of an frame have 
obviously different background levels. We identified these runs (and frames) 
by comparing the background levels between the two halves and scaling 
(multiplying) one side of the frame to match the other side. 
This has negligible impact on
the determination of extinction and the quality of final co-adds. The 
difference between the two halves is usually several DNs, and the fraction of 
affected frames is tiny ($\le2$\%, depending on scanline and filter). 

In the next step we rejected frames with very poor seeing, high sky 
background, or high extinction. We required that seeing should be better than
$2\farcs3$ and extinction should be smaller than 0.75 mag in the $r$ band
(Figure 4).
Although the cuts on seeing and extinction were made on the $r$-band frames, 
we rejected all associated data in the other four bands. However, unlike
\citet{ann11}, our selection cut on sky background was made on each band
separately, rather than on just the $r$ band. This is because the 
distributions of sky background for the five bands are very different, due to 
the fact that the dependence of sky background on is much more complex than 
the dependence of seeing or extinction on wavelength, especially under 
non-photometric observing conditions with significant moonlight or clouds.
Figure 4 shows the distributions of sky background in the five bands. The 
difference between the distributions is obvious. For example, the $u$ and $g$ 
band distributions have very long tails at the high background end, but the 
$z$-band distribution does not show a long tail. Therefore, we set different 
cuts on sky background for different bands. 
We rejected frames with background higher than [80, 250, 450, 550, 250] 
in units of DN in the five bands. This is the result of a tradeoff between
the background brightness and the number of frames to be rejected. 
The above cuts roughly correspond to [3.5, 3.5, 3, 2.5, 2.0] times the median 
background values in the five bands. The fractions of the frames rejected by 
the criteria are about [18\%, 21\%, 9\%, 4\%, 1\%], respectively. This means 
that our final co-adds include more $z$-band frames than $u$.

\begin{figure} %f4
%\epsscale{0.5}
\plotone{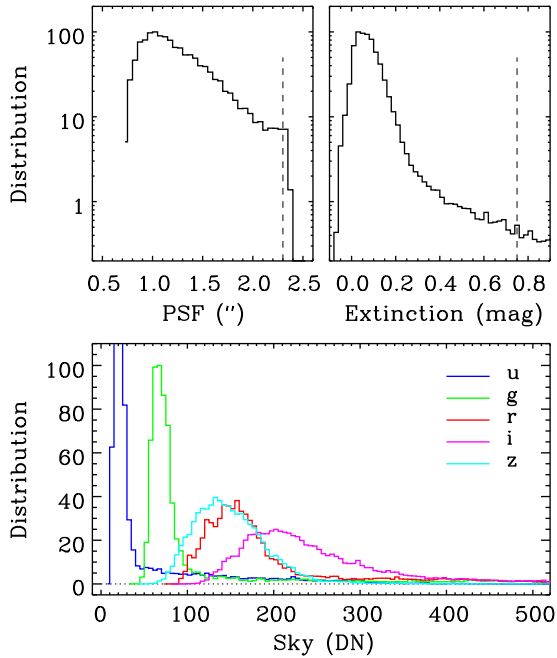}
\caption{Distributions of PSF, extinction, and sky background. The two
upper panels show the distributions of PSF and extinction measured in the
$r$ band. They have been normalized so that the peak values are 100.
The vertical dashed lines indicate our selection cuts. The lower panel shows
the distributions of sky background in the five bands. They have been
normalized so that the peak value in the $g$-band is 100. The distributions
are very different in different bands, so our selection cuts on sky background
were made on the individual bands, i.e., we rejected frames with background
higher than [80, 250, 450, 550, 250] in the five bands, respectively.}
\end{figure}

Our selection criteria above are considerably more permissive than those of
\citet{ann11}, in order to include as many frames as possible and enhance the 
depth. The marginal gain (in terms of the depth of the co-adds) by changing 
the values in the criteria (to include more input frames) is negligible. 

\subsection{Sky Subtraction}

The SDSS {\tt fpC} frames do not have the sky background removed. 
We used a simple but efficient method to perform sky subtraction.
Sky subtraction for the frames taken on moonless photometric nights are 
relatively easy, because the spatial and temporal variation across one SDSS
field is small. It becomes more straightforward if a field does not have any
large bright objects. In this case a low-order two dimensional function will 
provide a good fit to the background after objects are detected and masked 
out. However, a significant number of the runs were taken with moonlight and 
non-photometric conditions. They sometimes show strong background variation 
along the drift scan (R.A.) direction. Our background subtraction method was 
able to efficiently handle these extreme cases, as we now describe.

For each {\tt fpC} frame, we ran {\tt SExtractor} to detect objects 
after the soft bias was subtracted, and made a mask image accordingly. We set 
the detection threshold for objects to be $2\sigma$ in a minimum of 4 pixels.
We divided the frame into $12\times16$ grid elements, where each element 
measures 128 pixels at a side. We calculated a sky value for each grid element 
from $256\times256$ pixels centered on this element, after the masked pixels 
were removed. 
This sky value was computed based on the distribution of the pixel 
values. We first calculated the mean and standard deviation of the sky
pixels, which were then used to reject outliers. We repeated this process 
up to 20 iterations. If the mean was smaller than the median, then the mean
was adopted for the sky value; otherwise the sky value was computed by 
3 $\times$ median -- 2 $\times$ mean, an estimate for the mode of the 
distribution.
If more than half of the pixels in a 
grid element were masked out (usually due to the presence of very bright/large 
objects), this element was flagged as `bad', and its closest neighbors were 
also flagged as `bad'. The `bad' grid elements were not used. 

The grid elements that were not flagged as `bad' could still be affected by 
`bad' elements. We corrected for this using a simple method. Although the sky 
varies along the drift scan direction (rows, or the R.A. direction), it is 
usually stable in the perpendicular direction 
(columns, or the Decl. direction). 
A linear fit is a good description of the sky for each column for all but the 
worst data (which has already been rejected). However, if a grid element is 
affected by a very bright/large object, its sky value will deviate from the 
linear relation. We found the best linear relation for each column of 16 sky 
values (less than 16 if we already rejected some `bad' elements) in three 
iterations. In the first iteration we fit a simple linear relation to the data 
points. In the second iteration we fit a linear relation with a weight at each 
point. The weight at point $i$ is proportional to an exponential function 
$1/{\rm exp}(s_i-s'_i)$, where $s_i$ is the measured sky value and $s'_i$ is 
the value from the best linear fit in the first iteration. The exponential 
function strongly favors low values, which is consistent with the fact that 
lower values are almost always closer to real sky background. The third 
iteration repeated the algorithm from the second iteration, 
where the $s'_i$ value in the weight was from the second iteration.

\begin{figure*} %f5
%\epsscale{1.0}
\plottwo{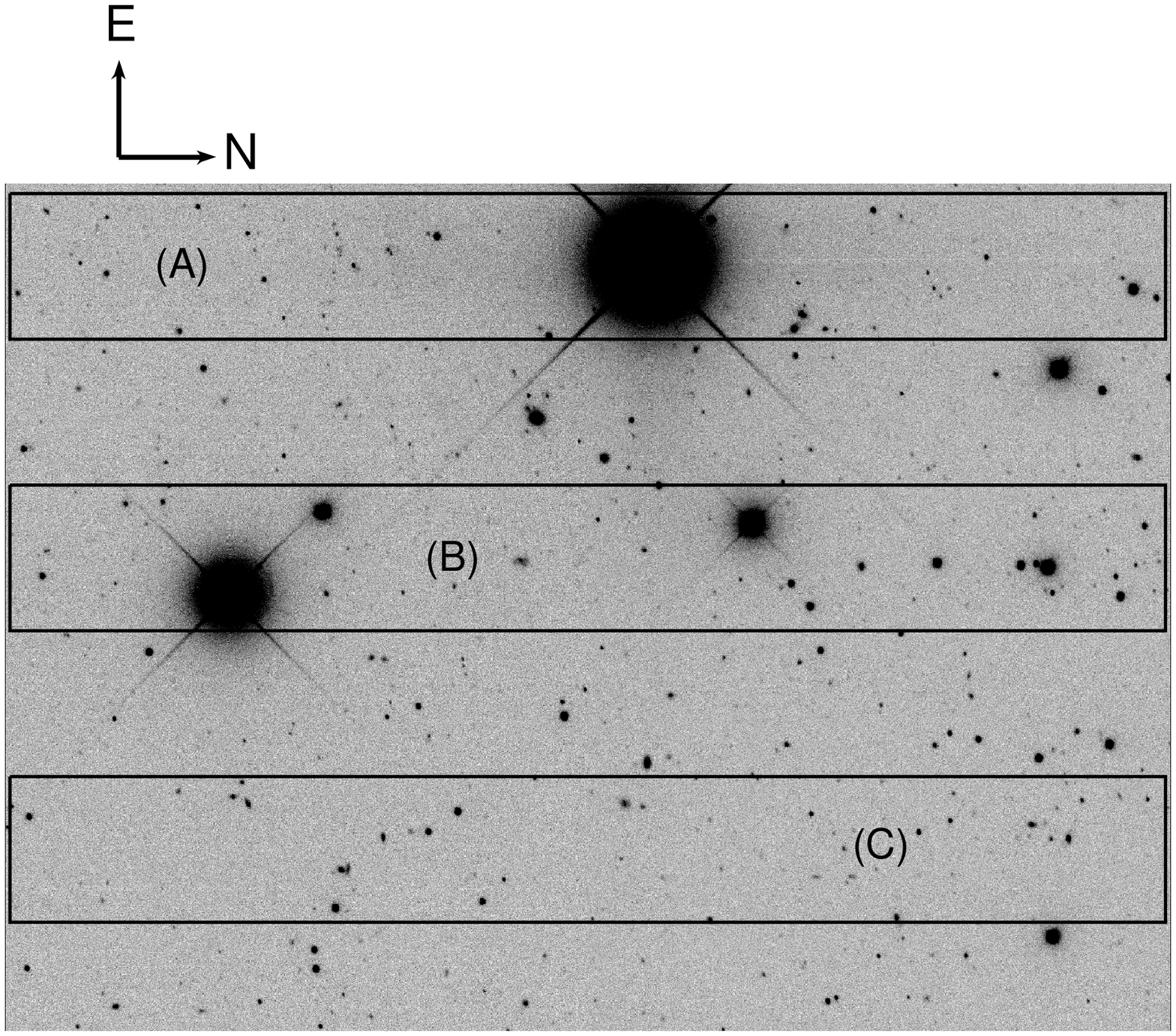}{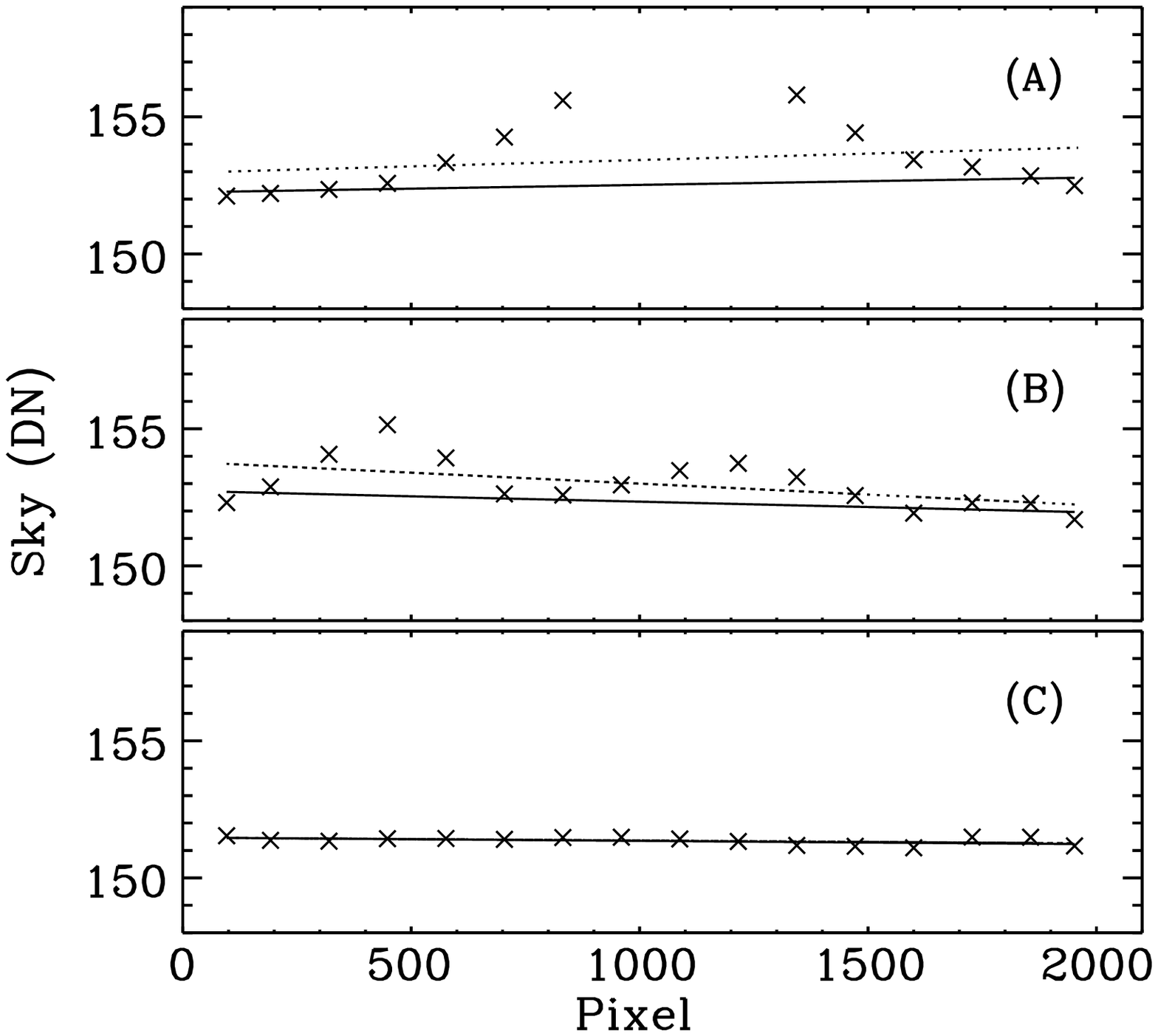}
\caption{An example of sky fitting. The left is an {\tt fpC} frame in the $i$
band, and the right panel demonstrates the sky fitting for the three columns
highlighted in the left image. The crosses represent sky values measured in
grids of $256\times256$ pixels, after the masked pixels are removed. The
dotted lines are the best linear fits from the first iteration of sky fitting.
The solid lines are the best fits from the third or last iteration, and
provide excellent estimates of real background. Note that for Column C where
there are no bright objects, the dotted and solid lines overlap.}
\end{figure*}

Figure 5 shows an example of sky fitting. The left is an {\tt fpC} frame in 
the $i$ band, with three columns (256 pixels wide) highlighted. The right plot 
shows the sky values (crosses) and their best linear fits (dotted and solid 
lines). In column A there is a large object, whose 3$\sigma$ detection area
extends more than $2\arcmin$ in diameter. Three grid elements of the sky have 
been flagged as `bad', and most elements in this column are affected by the 
presence of this object, as we can see from the big bump in the sky 
distribution. The dotted line is the best linear fit in the first iteration. 
The solid line is the best fit from the third iteration, and provides an 
excellent estimate of the real background. In column B, there are two bumps in 
the sky distribution, obviously due to the presence of the two bright objects. 
Although most of the grid elements in this column are affected by the two 
objects, our fitting process provides an excellent estimate of the sky.
Column C in the figure exhibits an ideal case in which there are no bright
stars in the whole column, so its 16 sky values can be well described by
a simple linear fit in the first iteration.

After each grid element was replaced by the best linear fit in the column
direction, the sky image was created by interpolating all pixels over the 
frame from the $12\times16$ sky values, using a minimum curvature spline 
surface. In order to save computing time, the spacing of the interpolation is 
$2\times2$ pixels, i.e., the resolution in the final sky image is $2\times2$ 
pixels ($0\farcs8 \times 0\farcs8$).

The sky subtraction algorithms used by the SDSS pipeline {\tt PHOTO} 
\citep{lup01} are
similar to the first part of our method. {\tt PHOTO} first detects and masks
out bright objects, and computes sky values in grids of $256\times256$ pixels.
These sky values are linearly interpolated to make a sky image, without 
further treatment of outliers. This could systematically underestimate the 
brightness of large objects. As shown in Figure 5, the bright object in column
A is very extended. If we simply use the measured sky values (crosses), the 
sky around the object will be significantly overestimated, so that the 
brightness of the object is underestimated. Our fitting procedure can
properly estimate the sky in the vicinity of this bright object.
Our tests show that we can compute reasonable sky values unless objects are
so large that they affect all sky grid elements in a column. Even in this 
case, our algorithm can still minimize the effect from bright objects.

\citet{ann11} used a different method to estimate sky background. 
They computed a median value for each column of pixels, and then fitted a 
linear relation to these median values along the row direction with sigma
clipping in five iterations. This approach assumes that the sky is constant
along the Decl. direction and varies linearly along the R.A. direction.
This is usually overly simplistic for images taken with significant moonlight 
or clouds. Our method is more sophisticated, by assuming a linear relation
along the Decl. direction and by allowing arbitrary sky variation along the
R.A. direction.

\subsection{Image Weights}

Before we co-added the {\tt fpC} frames, each input frame was assigned a 
weight proportional to
\begin{equation}
\frac{T^{k}}{{\rm FWHM}^m\ \sigma^n}\ ,
\end{equation}
where $T$ is the sky transparency as measured by the extinction of the frame, 
FWHM is the FWHM of the PSF, $\sigma$ is the standard deviation of the 
background noise (one constant value for a frame), and $k,m,n$ are indices 
to be set for a specific science goal. In order to maximize the 
signal-to-noise ratios (S/Ns) of the sources in the co-added images, we set 
$k=1$ and $n=2$. In the $griz$ bands, the noise in frames is completely 
dominated by sky background, so $\sigma^2$ is the variance of the sky 
background. In some $u$-band frames with very low sky background, however, 
read noise is not negligible. So we added read noise in quadrature
to $\sigma^2$ in the $u$ band.

For FWHM, it is difficult to determine an optimal $m$ value for our co-adds. 
For point sources in individual frames in which sky background dominates the 
noise, the optimal $m$ value is 2 (thus the depth is proportional to 
1/FWHM$^2$), because the area occupied by an unresolved object scales with 
FWHM$^2$. However, for a co-added image with a large number of input frames 
and a wide range of PSF FWHM, the optimal $m$ is less obvious. We carried out
simulations based on artificial images and tests based on real Stripe 82 
images, which showed that the S/N of the sources in co-adds is very 
insensitive to $m$ in the range $0.5<m<2$. Considering that the PSF FWHM is 
less important for more extended sources, we chose $m=1$ for FWHM instead of 
$m=2$. In summary, the indices we chose are $k=1$, $m=1$, and $n=2$, thus the
weight is proportional to $T/({\rm FWHM}\ \sigma^2)$. This is different from
the weight ($T/({\rm FWHM}^2\ \sigma^2)$) used by \citet{ann11}.

For each input frame, we read the list of defective pixels, i.e., those 
affected by cosmic rays and bad columns, from the associated {\tt fpM} file,
and assigned them a near-zero weight (1.0$^{-10}$). We also assigned this 
near-zero weight to the overlap region (128 pixel wide) between one frame 
and its following neighbor frame. After that, a weight image was created for 
each input frame.

\subsection{Co-addition}

We divided each scanline along $\rm -60\degr<R.A.<60\degr$ into 401 regions 
each with a size of $2850\times2048$ pixels. The pixel size is the same as the 
native size of $0\farcs396$. There is also a common area of 128 pixels between 
each region and its following neighbor region. The size of the regions is 
twice the size of the SDSS {\tt fpC} frames when the overlap area is removed. 
This reduces the total number of output co-added images by a factor of two, 
but still allows one to easily split one co-added image into two images with 
the {\tt fpC} frame size. As we described in Section 2.1, we rejected frames 
based on PSF, extinction, and sky background. At this stage, we made a final 
selection cut for each region in each band, removing any frames with weights 
below 0.4 times the median weight of the frames belonging to that region. 
This removed $\le2$\% of frames.

With the corrected input {\tt fpC} frames and their weight maps, the
construction of co-adds is straightforward. We scaled the frames by $1/T$, and 
re-sampled them to a common astrometric grid using {\tt SWARP} \citep{ber02}. 
Note that reliable astrometric solutions (better than 45 mas rms; Pier et al. 
2003) have been incorporated into SDSS {\tt fpC} frames. The re-sampling 
interpolations for the science and weight images were lanczos3 and bilinear, 
respectively. We then co-added images using {\tt SWARP}. The co-addition is a 
weighted mean with outlier rejection ($7\sigma$). The output products include 
a co-added science image and its associated weight image for each region. The 
weight images record relative weight at each pixel position. 

The main differences between our co-adds and those of \citet{ann11} are as
follows. First, our co-adds include many more SDSS runs, but our selection of
runs and fields is more permissive. Second, the sky subtraction algorithms
and image weights are slight different. Finally, \citet{ann11} produced
catalogs of objects with the SDSS pipeline, but we produced object catalogs
with {\tt SExtractor} (Section 3.2).

\section{DATA PRODUCTS}

Because of our selection cuts on PSF, extinction, and sky background, the
actual number of SDSS frames used in the co-adds varies from region to region.
In Figure 2, the blue (red) dashed and dotted lines show the number of frames
used for the co-adds of N (S) scanline 2 in the $i$ and $g$ bands, 
respectively. The number of frames used for $i$ is 60--70, and for $g$ is
50--60. These numbers are more than twice larger than those used in 
\citet{ann11}. For each filter, the fraction of frames used in our co-adds is 
roughly constant with right ascension. For example, this fraction is 
$\sim90$\% in the $i$ band and $\sim75$\% in the $g$ band. In this section, 
we will introduce our final data products and provide some basic statistical 
information for the co-adds.

\subsection{Photometric Calibration}

Although atmospheric extinction was measured and corrected for individual 
input frames, it was not very accurate for frames with large extinction and/or 
high sky background due to the small numbers of high S/N point sources 
(especially in the $u$ and $z$ bands). So we determined photometric 
calibration on the co-added images. As we did in Section 2.1, we ran 
{\tt SExtractor} on each co-added image and performed aperture photometry 
for point sources within an aperture (diameter) size of 20 pixels 
($8\arcsec$). Blended objects were rejected. We then matched this object 
catalog to the \citet{ive07} catalog, and computed a zero point (the median 
magnitude difference of the matched objects between the two catalogs is zero). 
For simplicity, we already assumed the exposure time of 1 sec for the zero 
points, so the magnitude of an object is simply --2.5 log(DN) + zero point.
For the region of sky that the \citet{ive07} catalog does not cover, we 
applied an average zero point for each scanline. 
The zero points were recorded in the image headers as `magzero'.

\begin{figure} %f6
%\epsscale{0.5}
\plotone{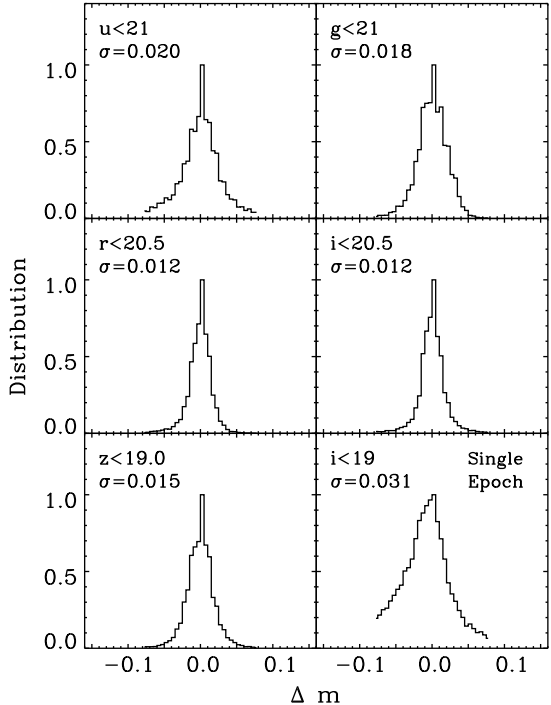}
\caption{Distributions of the differences between the aperture ($4\arcsec$
radius) magnitudes of the co-added images and the PSF magnitudes from the
\citet{ive07} catalog ($\rm R.A.=300\degr - 30\degr$; scanline 08).
The distributions have been normalized so that the peak values are 1. The
$\sigma$ values are from the best Gaussian fits. The right bottom panel shows
the difference between the aperture ($7\farcs4$ radius) and PSF magnitudes for
bright ($i<19$) stars from SDSS single-epoch data.}
\end{figure}

In Figure 6 we show the distributions of the magnitude differences between
the aperture magnitudes of the co-added images and the PSF magnitudes from
the \citet{ive07} catalog for a subset of our data 
($\rm R.A.=330\degr - 30\degr$; scanline 08). 
The $\sigma$ values are from the best Gaussian fits, and have a range from 
0.012 to 0.020. These magnitude differences are likely caused by the 
combination of the different algorithms use to measure aperture and PSF 
magnitudes and the photometric errors or calibration. For comparison, 
in the right bottom panel we show the distribution of the difference between 
the aperture ($7\farcs4$ radius) and PSF magnitudes for bright ($i<19$) stars 
from SDSS single-epoch data. Its $\sigma$ value is 0.031.

\subsection{Images and Catalogs}

Our final products consist of 24060 science images (12 scanlines $\times$ 5
filters $\times$ 401 regions) and their associated weight images and catalogs. 
The science and weight images are named `S82\_xxy\_zzz.fits' and
`S82\_xxy\_zzz.wht.fits', where `xx' is the scanline number from 01 to 12
(Figure 1), `y' is the filter, and `zzz' is the region number from 001 to 401.
For example, `S82\_08i\_234.fits' and `S82\_08i\_234.wht.fits' are the co-adds 
for scanline 08 (N scanline 2), filter $i$, and region 234. 

We also produced object catalogs from these data. The catalogs are named 
`S82\_xxy\_zzz.cat', and were produced by {\tt SExtractor}. The key part of 
the {\tt SExtractor} configuration file is displayed in Table 1. Briefly, we 
detected objects in a minimum of 4 contiguous pixels (DETECT\_MINAREA) with a 
detection threshold of $2\sigma$ (DETECT\_THRESH). {\tt SExtractor} took the 
weight images produced by {\tt SWARP} as input weight images (WEIGHT\_TYPE) 
during its procedure of object detection. {\tt SExtractor} deblending is done 
using a multi-thresholding algorithm. It deblends components of a composite 
detection at up to DEBLEND\_NTHRESH levels based on local detection peaks, 
where DEBLEND\_NTHRESH is the number of deblending sub-thresholds. At each 
level, DEBLEND\_MINCONT is the minimum contrast parameter for deblending, 
i.e., any new component with flux larger than DEBLEND\_MINCONT (times the 
total flux of the current `parent' component) is considered as a new component 
for the next level of deblending.
We performed aperture photometry within five diameter aperture 
(PHOT\_APERTURES) sizes of [6, 8, 10, 12, 20] pixels ($2\farcs4$, $3\farcs2$, 
$4\farcs0$, $4\farcs8$, and $8\farcs0$,). We also computed `AUTO' magnitudes 
(Kron-like elliptical aperture magnitudes; the best for extended sources from 
{\tt SExtractor}) and Petrosian magnitudes with the default parameters 
(PHOT\_AUTOPARAMS and PHOT\_PETROPARAMS). 

\begin{figure} %f7
%\epsscale{0.5}
\plotone{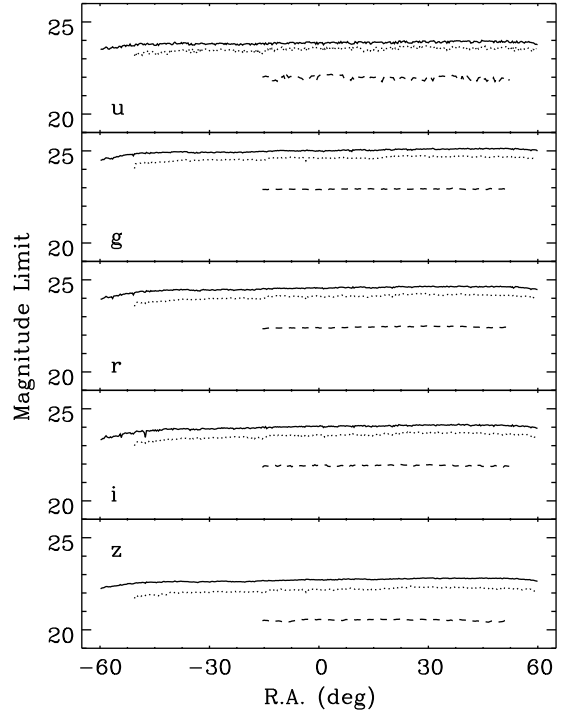}
\caption{The $5\sigma$ detection limits of the aperture ($3\farcs2$ diameter)
magnitudes for point sources in scanline 08 of the co-adds (solid lines).
The dotted lines are the magnitude limits for the \citet{ann11} co-adds, and
the dashed lines are the magnitude limits for single-epoch (Run 4263) data.
Run 4263 is one of the best runs for Stripe 82. Our co-adds are 1.9--2.2 mag
deeper than the best SDSS single-epoch data, and 0.3--0.5 mag deeper than the
\citet{ann11} co-adds.}
\end{figure}

\begin{figure*} %f8
%\epsscale{0.8}
\plotone{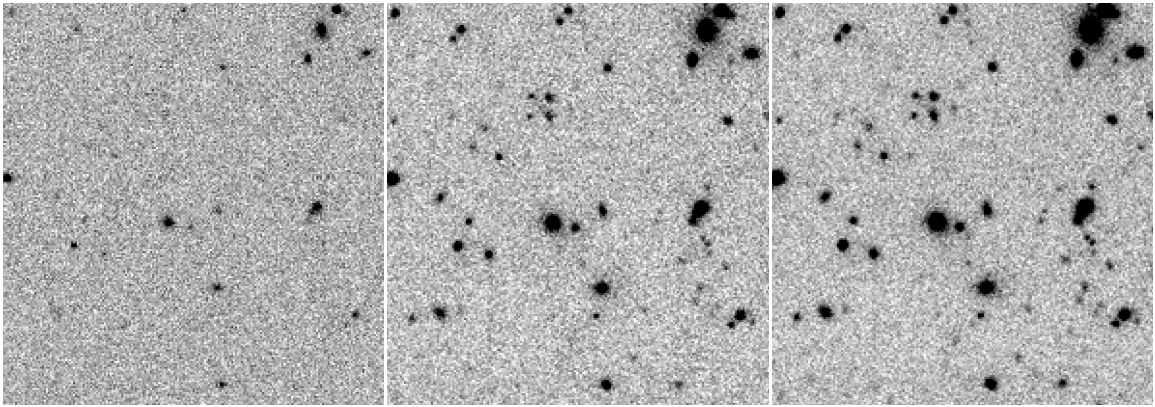}
\caption{Direct comparison between a single-epoch $i$-band frame (left panel),
the \citet{ann11} co-add (middle panel), and our co-add (right panel).
The PSF sizes are similar ($0\farcs8$--$0\farcs9$).
The image size is $1.3\times1.3$ arcmin located at 21h29m30s--00d32m20s.}
\end{figure*}

The complete list of the quantities computed for the catalogs is shown in 
Table 2. Each catalog can be roughly divided into three parts. The first part 
shows the positions and coordinates of detected objects. The second part lists
various magnitudes and errors. The third part provides some basic information
about object structure and morphology, including semi-major and minor axes,
ellipticity, FWHM, etc. The last parameter `FLAGS' is the {\tt SExtractor} 
extraction flag, the sum of powers of 2. For example, `1' ($2^0$) means that 
an object is very close to bright objects or bad pixels so that its photometry 
could be significantly affected. `2' ($2^1$) means that an object is blended
with other objects. The detailed explanations for the keywords and 
parameters in Tables 1 and 2 can be found in \citet{ber96} and the 
{\tt SExtractor} user's manual. 
There are several caveats on how to use these catalogs.
\begin{enumerate}
\item There is overlap between one region and its following neighbor 
region, so any object detected in this overlap area will show up in two
catalogs. In addition, any two adjacent scanlines also slightly overlap,
and these are independent detections.
\item The catalogs are not matched between the bands, resulting in different
object lists for each band within the same region.
The deblending of an object in different bands could also be different.
\item The object detection threshold is $2\sigma$ in a minimum of 4 pixels, 
so objects fainter than this are not included. One example is very low surface
brightness galaxies. In order to detect such galaxies, one may convolve the
images with a kernel.
\item Users need to apply aperture corrections before aperture magnitudes 
can be used across different bands (e.g. for constructing color-color 
diagrams). Aperture 
corrections are different for different regions in different bands (depending 
on wavelength and PSF). A good approximation for point sources within a given 
aperture is the median difference between the aperture magnitudes within this 
aperture and the aperture magnitudes within 20 pixels for bright isolated 
point sources from the catalogs (aperture corrections from 20 pixels to
infinite are smaller than 0.5\%).
\item Measurements of objects that are several pixels away from image edges 
are not reliable. Because these objects are in the overlap area between 
adjacent regions or scanlines, better measurements can be found in the 
adjacent catalog where the object is less near to the edge. 
\item Transient objects and objects with high proper motions have been
eliminated by the use of outlier rejection when co-adding the images.
Also note that many single-band detections are spurious detections, usually
associated with diffraction and bleed spikes from bright stars. These spikes
can be faint in the co-adds (invisible in input SDSS frames), but extend 
across more than one image. 
\item Our catalogs only included commonly-used quantities, usually with
default setup parameters. If users need more quantities, or quantities with 
different parameters, they are advised to run {\tt SExtractor} (or other
tools) on the co-added images by themselves.
\end{enumerate}

\subsection{Depth}

\begin{figure} %f9
%\epsscale{0.5}
\plotone{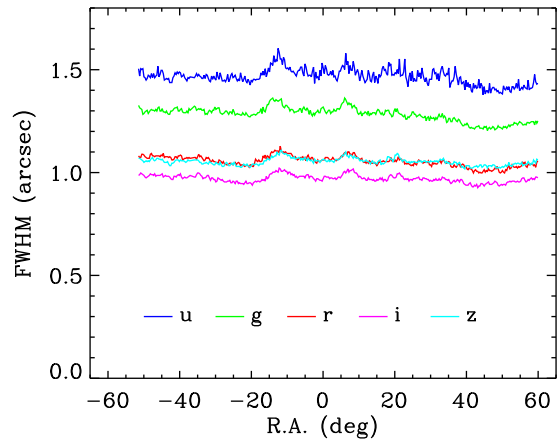}
\caption{PSF FWHM in the five bands in scanline 08 (N scanline 2) of
the co-adds. The PSF FWHM in the $riz$ bands is roughly $1\arcsec$, and in the
$ug$ bands is about $1\farcs3-1\farcs5$.}
\end{figure}

We determine the depth of our co-added data based on the aperture magnitudes
within an aperture (diameter) size of 8 pixels ($3\farcs2$). As the 
photometric uncertainties depend on aperture size, our choice of 8 pixels, or 
2--3 times the PSF FWHM, represents a tradeoff between reducing the aperture 
correction and reducing background noise. In Figure 7 we show the magnitude 
limits of the $5\sigma$ detection (i.e., photometric errors = 0.22) for point 
sources in scanline 08 (solid lines). Our co-added images have great depth of
roughly 23.9, 25.1, 24.6, 24.1, and 22.8 AB magnitudes in the five bands, 
respectively. The magnitude limits are compared to the magnitude limits of the 
\citet{ann11} co-adds (dotted lines) and of the single-epoch (Run 4263) data 
(dashed lines). Run 4263 is one of the best runs for Stripe 82, taken on a 
photometric dark night with excellent seeing ($\rm PSF\sim0\farcs8$ in the $r$ 
band). The magnitude limits for all three different datasets were computed 
using the same method, i.e., aperture photometry within an aperture of 8 
pixels. Compared to Run 4263, our co-adds are 1.9, 2.1, 2.1, 2.1, and 2.2 mag 
deeper in the $ugriz$ bands, respectively. Our co-adds are also 0.3--0.5 mag 
deeper than the co-adds of \citet{ann11}.
%Quantitatively, our co-adds 
%are 0.3, 0.4, 0.5, 0.5, and 0.5 mag deeper in the five bands, respectively.
Figure 8 shows a direct comparison between a single-epoch frame in the $i$ 
band (left), the \citet{ann11} co-add (middle), and our co-add (right).

Figure 9 shows the PSF FWHM in the five bands in scanline 08 of the co-adds.
The PSF FWHM was measured with {\tt SExtractor} based on bright point sources.
Note that PSF measurements from different methods (such as SDSS,
{\tt SExtractor}, or {\tt IRAF}) can be slightly different due to the
different algorithms used. The $i$-band images in the co-adds have the PSF
with the smallest FWHM, and the $u$-band images have the worst PSF. The $r$
and $z$ band images have similar PSF sizes. In addition, the PSF varies across
scanlines due to the camera optics. SDSS scanline 6 (scanlines 06 and 12 in
our co-adds) has the worst PSF in most bands. Figure 9 shows that the PSF FWHM
in the $riz$ bands is roughly $1\arcsec$, and in the $ug$ bands is between
$1\farcs3$ and $1\farcs5$. These numbers are consistent with those of
single-epoch data with the best observing conditions, and are also consistent
with the co-adds of \citet{ann11}.

\subsection{Color-Color Diagrams}

SDSS point sources are mostly main sequence stars, which form a tight stellar 
locus in color-color diagrams \citep{ive04}. The width of the stellar locus is 
almost independent of magnitude, but is broadened by photometric errors of 
stars, so color-color diagrams are a useful tool for photometric quality 
assessment. We first separate stars (point sources) from galaxies (extended
sources). The SDSS uses the difference between PSF magnitude and so called
`model' magnitude to do star-galaxy separation. As we did not run the SDSS
pipeline, we did not measure these magnitudes. Instead, we separate stars and
galaxies based on the distribution of object sizes (FWHM) as measured by 
{\tt SExtractor}. Figure 10 shows an example. The upper panel shows the object 
size (FWHM) as the function of the brightness in $r$ 
($\rm R.A.=10\degr-30\degr$; scanline 08). The object detection and photometry 
are described in Section 3.3. The narrow horizontal band in the plot clearly 
demonstrates the location of stars. The width of this band is dominated by
the small variation of the PSF across the scanline.
The lower panel shows the distribution of the object sizes. At $r<22$ mag,
stars in this example can be well separated using $\rm FWHM<1\farcs12$. They 
start to mix with galaxies at $r>22$ mag, and are completely mixed with 
galaxies at $r>23$ mag, as seen in the both panels. But we still use 
$\rm FWHM<1\farcs12$ to separate stars from galaxies at $r>23$ mag. 

\begin{figure} %f10
%\epsscale{0.5}
\plotone{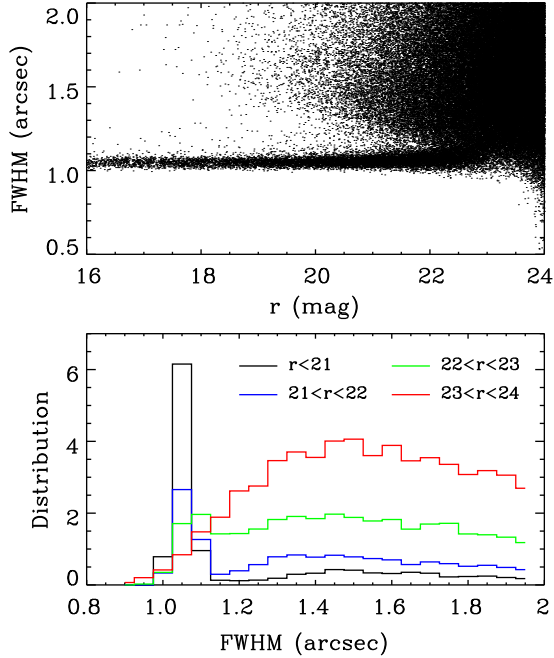}
\caption{Star galaxy separation. The upper panel shows the object size (FWHM)
as a function of the brightness in $r$ ($\rm R.A.=10^h-30^h$; scanline 08).
The narrow strip clearly indicates the location of stars. The lower panel
shows the FWHM distribution. At $r<22$ mag, stars are well separated. They
start to mix with galaxies at $r>22$ mag, and are completely mixed with
galaxies at $r>23$ mag. In this example we use $\rm FWHM<1\farcs12$ to
separate stars from galaxies.}
\end{figure}

In Figure 11 we show the $r-i$ versus $g-r$ color-color diagram for point 
objects brighter than $r=24$ mag selected in Figure 10. The objects are 
grouped into four magnitude bins. As expected, the stellar locus in the 
brightest bin ($r<21$ mag) is very tight. It becomes broader in fainter bins, 
as photometric errors start to dominate the width. In the third bin 
($22<r<23$), the stellar locus still has a well-defined shape, though it
is much broader. In the faintest bin, the stellar locus is not as obvious as
it appears in the brighter bins, due to large photometric errors and leakage 
of a large number of galaxies, since the star-galaxy separation does not work 
well at the faintest end (Figure 10).

\begin{figure} %f11
%\epsscale{0.5}
\plotone{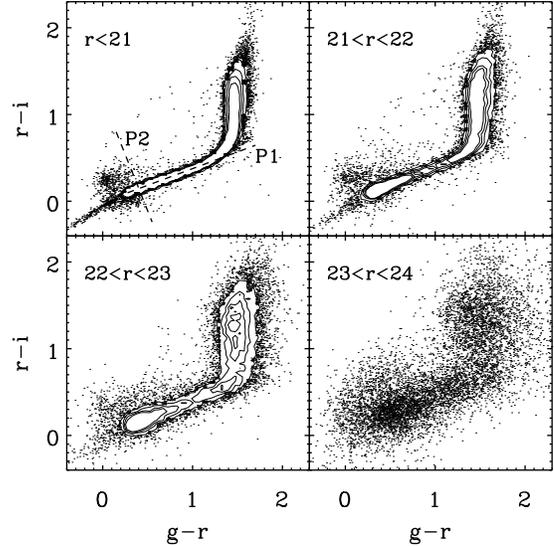}
\caption{The $r-i$ versus $g-r$ color-color diagram for point objects brighter
than $r=24$ mag selected in Figure 10. In the first panel we also define
two principal axes, P1 and P2. P1 is along the blue part of the stellar locus
and P2 is perpendicular to P1. They are used to quantify the width of the
stellar locus (see Section 3.4 and Figure 12).}
\end{figure}

We quantify the width of the stellar locus in Figure 11 following the method 
of \citet{hel03} and \citet{ive04}. We focus on the width of the blue part
in the $r-i$ versus $g-r$ diagram, and define two principal axes, P1 and P2.
As shown in the first panel of Figure 11, P1 is along the locus and P2 is
perpendicular to P1. P2 is further adjusted for a weak dependence on $r$.
The $w$ color is then defined on P1 and P2 as the distance from a star to P1.
The distribution of $w$ describes the width of the stellar locus. The results
are shown in Figure 12, where the stars brighter than $r=23$ mag from 
Figure 11 are grouped into four bins. The $\sigma$ values are from the best 
Gaussian fits, and have a range from 0.016 (brightest bin) to 0.061 (faintest
bin). \citet{ive04} reported that the rms of $w$ at $r<20$ mag is 0.025 mag 
for SDSS single-epoch data, and decreases to 0.022 mag for data with 
observations at several epochs.
We reached $\sigma=0.019$ mag at $r<22$ mag, indicating that our co-adds are 
indeed at least two mag deeper than single-epoch data. \citet{ive07} also 
reported a rms of $w$ of $\sigma=0.010$ mag at $r<20$ mag for 
data with multi-epoch ($\ge10$) observations. We achieved 0.016 mag at
$r<21$ mag and could not obtain a smaller $\sigma$ for brighter stars, 
suggesting the existence of a calibration floor in the data.
This was likely caused by the difference between the aperture magnitudes of
the co-adds and the PSF magnitudes of the \citet{ive04} catalog during the
process of photometric calibration (Section 3.1 and Figure 6).

\section{NOAO/NEWFIRM $J$-BAND IMAGING DATA}

In addition to the SDSS data in five optical bands, Stripe 82 is also 
(partially) covered by surveys/observations at many other wavelengths, such as 
the UKIRT Infrared Deep Sky Survey \citep[UKIDSS;][]{law07}, the Very Large 
Array imaging of Stripe 82 at 1.4 GHz \citep{hod11}, and the Herschel Stripe 
82 Survey \citep{vie13}. In this section we present our near-IR $J$-band 
observations with NOAO NEWFIRM \citep{pro04}. NEWFIRM is a wide field infrared 
imager with a field-of-view of $27\arcmin\times27\arcmin$ (pixel size 
$0\farcs4$), mounted at the f/8 R-C focus of the NOAO 4-m telescopes. Our 
NEWFIRM observations were made with the Kitt Peak 4m Mayall telescope. They 
cover $\sim$90 deg$^2$ of Stripe 82 to a depth of 20--20.5 Vega mag.

\begin{figure} %f12
%\epsscale{0.5}
\plotone{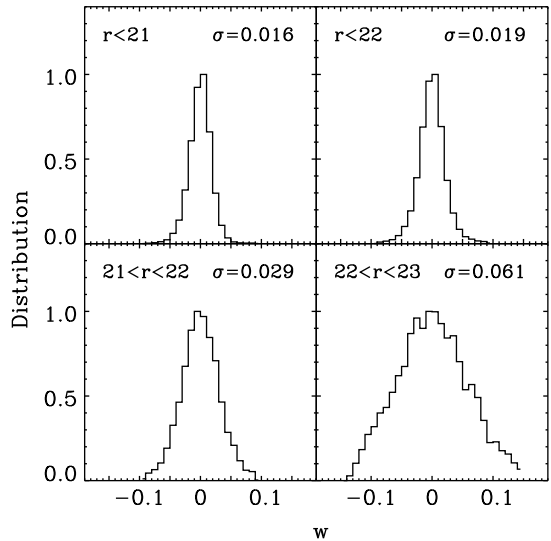}
\caption{Distributions of the $w$ colors for the stars brighter than $r=23$
mag in Figure 11. The distributions have been normalized so that the peak
values are 1. The $\sigma$ values are from the best Gaussian fits.
The value of $\sigma$ (0.019 mag) at $r<22$ is smaller than the rms of $w$ at
$r<20$ for SDSS single-epoch data, indicating that the co-adds are at least
two mag deeper than single-epoch data.}
\end{figure}

\subsection{Observations and Data Reduction}

The NEWFIRM observations were made in two runs on 2007 November 10--16 and 
2009 January 6--15. The observing conditions in the two runs were moderate,
with mostly clear skies and a large range of seeing from $\sim1\arcsec$ to 
$>2\farcs5$. In the 2007 run, we observed $\sim$150 fields (NEWFIRM 
pointings). A typical observing strategy was a $3\times3$ dither pattern
(dither offset $40\arcsec$). The exposure time at each dither position was 30 
sec or 60 sec, depending on sky background. This pattern was conducted twice 
with slightly different central positions for data taken with an exposure time 
of 30 sec. 
The total integration time was thus 540 sec per field. In the 2009 run, we 
observed $\sim$300 fields. We used a dither pattern with five positions
(dither offset $45\arcsec$). At each dither position we took six short (15 
sec) exposures. The short exposures were co-added internally and read out as 
one image. The total integration time per field was also 540 sec. Note that 
this ability to do internal co-addition had not been embedded in the NEWFIRM 
observing pipeline during the 2007 run. Adjacent fields slightly overlapped
by $1\arcmin-2\arcmin$. Several fields were observed twice due to low image 
quality. We also rejected a small fraction of images that were taken with 
very poor observing conditions.

The NEWFIRM data were reduced with the combination of our {\tt IDL} routines
and the IRAF\footnote{IRAF
is distributed by the National Optical Astronomy Observatory, which is 
operated by the Association of Universities for Research in Astronomy (AURA) 
under cooperative agreement with the National Science Foundation.}
NEWFIRM task by M. Dickinson and F. Valdes. The basic procedure is summarized 
as follows. We first reduced calibration data, and made master dark and dome
flat images for each night. Each science image was then trimmed and a dark 
frame was subtracted, followed by linearity correction and flat fielding. A 
weight image was created by assigning a near-zero number ($10^{-10}$) to 
defective pixels, such as bad pixels, saturated pixels, and persistence. 
The weight image did not include seeing or sky transparency. Unlike the Stripe 
82 images that were taken under very different observing conditions, the 
NEWFIRM images for any single field were taken under similar conditions within 
a span of $\sim$10 min.
Sky subtraction was done using a similar method to the one we used for the 
SDSS images. After sky background was subtracted, we detected objects using 
{\tt SExtractor}, and calculated astrometric solutions using {\tt SCAMP} 
\citep{ber06} by matching objects to the SDSS. Finally we used {\tt SWARP} to 
re-sample and stack images, as we did for the SDSS images. The re-sampling 
interpolations for science and weight images were lanczos3 and bilinear,
respectively. The co-addition is a weighted mean with outlier rejection 
($5\sigma$). The products are one co-added science image and its associated
weight image for each field. The weight image records relative weight at each 
pixel position.

\begin{figure} %f13
%\epsscale{0.5}
\plotone{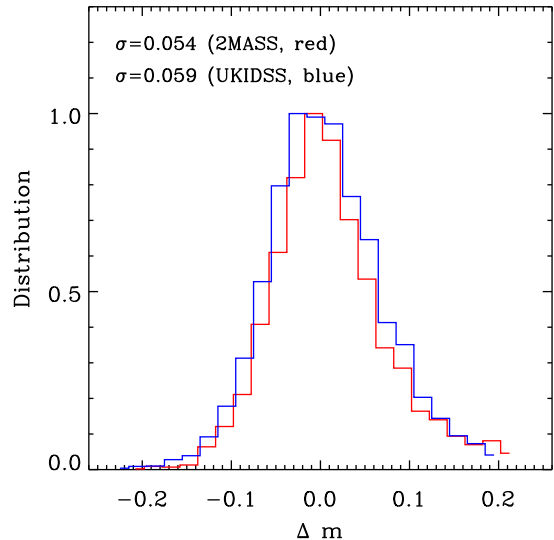}
\caption{Quality of the photometric calibration for the NEWFIRM data.
The histograms show the comparison of the calibrated NEWFIRM data with
2MASS (red) and UKIDSS (blue). The 2MASS objects are point sources brighter
than $J=15.5$ mag, and the UKIDSS objects are point sources between $J=17$ and
18 mag. The $\sigma$ values (rms of the dstributions) are from the best
Gaussian fits. The figure indicates that our photometric calibration is
accurate to about 5--6\%}
\end{figure}

\subsection{Photometric Calibration and Data Products}

We performed photometric calibration using the method that we did for the SDSS
co-adds. Briefly, we ran aperture photometry within an aperture (diameter) 
size of 20 pixels ($8\arcsec$). Blended objects were rejected. We then matched 
to the 2MASS \citep{skr06} point source catalog, and computed a zero point for 
each image. For simplicity, we already assumed the exposure time of 1 sec 
for the zero point, so the magnitude of an object is simply --2.5 log(DN) + 
zero point. The zero point was recorded in the image headers as `magzero'.

Figure 13 compares the calibrated NEWFIRM data and the 
2MASS point source catalog (red histogram). The NEWFIRM data displayed in 
this figure were taken on one 2007 night and one 2009 night. The 2MASS objects 
were chosen to be brighter than $J=15.5$ mag. The rms of the magnitude 
difference is $\sigma=0.054$,
meaning that our photometric calibration is accurate to about 5\%.
We also compared the NEWFIRM data with the UKIDSS data (blue histogram). 
The UKIDSS objects were chosen to be point sources between $J=17$ and 18 mag.
The distribution of the magnitude difference is consistent with the red
histogram. Note that UKIDSS is roughly three magnitudes deeper than 2MASS, 
and was also calibrated using the 2MASS catalog.

\begin{figure} %f14
%\epsscale{0.5}
\plotone{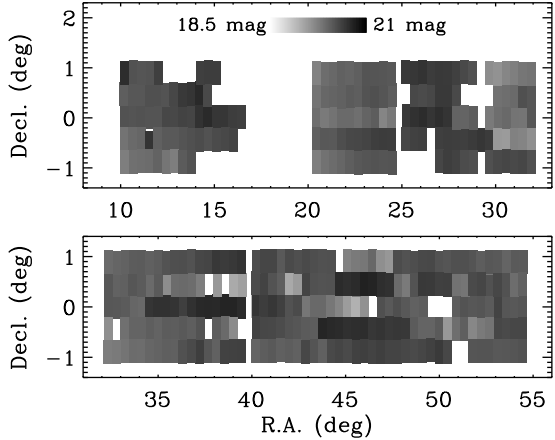}
\caption{Layout of the NEWFIRM pointings and the depth of the $J$-band images.
The depth is the $5\sigma$ detection limit for point sources.}
\end{figure}

\begin{figure} %f15
%\epsscale{0.5}
\plotone{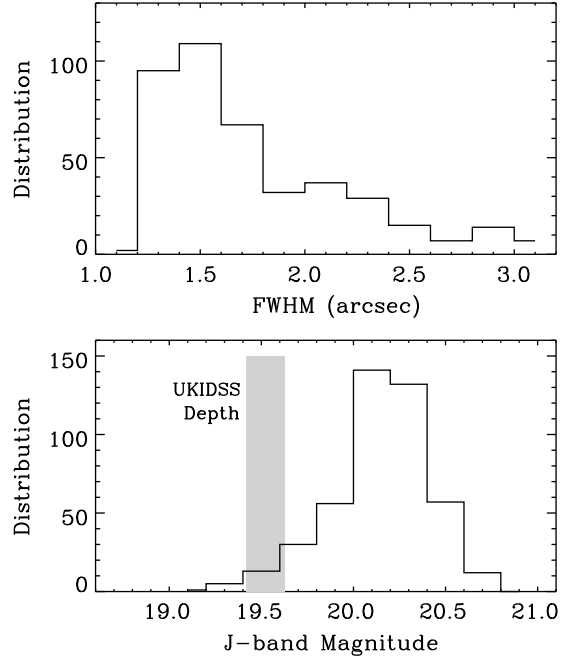}
\caption{Upper panel: Distribution of PSF FWHM in our NEWFIRM $J$-band images.
Lower panel: Distribution of the image depth for point sources. The grey
shaded region shows the UKIDSS $J$-band depth in Stripe 82 (see details in
Section 4.3). Our NEWFIRM images are about 0.7 mag deeper on average.}
\end{figure}

The final data products include 450 stacked images and the associated weight 
images and catalogs. The image size is 4300 by 4300 pixels, and the pixel size 
is the same as the native size of $0\farcs4$. The area at image edges 
($\sim$100 pixels) have much lower coverage due to the dithering of the
observations. When this is taken into account, the effective area of each 
image is about 0.2 deg$^2$. So our final products cover roughly 90 deg$^2$ of 
Stripe 82, in the range of $\rm 10\degr<R.A.<55\degr$.
Figure 14 shows the layout of the NEWFIRM pointings along with the 
image depth (the next subsection).
The catalogs were produced in the same way as we did for the SDSS images
(Section 3.2). The {\tt SExtractor} configuration and parameter files are
also the same as shown in Tables 1 and 2, except the aperture sizes.
We used the five sizes of [8, 10, 12, 14, 20] pixels for aperture magnitudes.

\subsection{Depth and Color-Color Diagrams}

We measure the depth of the images in the same way as we did for the SDSS
data. The depth is described as the $5\sigma$ detection limit for point 
sources. The photometry was measured in one of four apertures (diameter) 
[8, 10, 12, 14] pixels ($3\farcs2 - 5\farcs6$), because of the range of 
image quality in these data. The upper panel in Figure 15 shows the 
distribution of the PSF FWHMs. Many PSF FWHMs are larger than $2\arcsec$ due 
to poor seeing and unstable instrument focus in 2007. The lower panel in 
Figure 15 shows the distribution of the image depth. While the distribution 
spans a wide range from 19 to 21 mag, most images have a depth of 20--20.5 
mag. For comparison, the grey shaded region shows the UKIDSS depth (single 
epoch) in the $J$ band in Stripe 82. The depth is also the $5\sigma$ detection 
limit, derived from $\sim$100,000 point sources centered at $\rm R.A.=$ 
10$\degr$, 20$\degr$, 30$\degr$, and 40$\degr$ ($\rm Decl.=0\degr$). The left 
and right boundaries of the shaded region indicate the $1\sigma$ range of the 
detection limit distribution. On average, our $J$-band images are $\sim$0.7 
mag deeper than the UKIDSS images.

Compared to the SDSS co-added images, however, the $J$-band images are
significantly shallower (Figure 16).
The depth of the SDSS co-adds shown in Figure 7 is
24--25 mag in $ugri$, and is $\sim$22.8 mag in $z$. The $J$-band depth is
20--20.5 Vega mag, or 21--21.5 AB mag. This is 1.5$\sim$3 mag shallower.

\begin{figure} %f16
%\epsscale{0.8}
\plotone{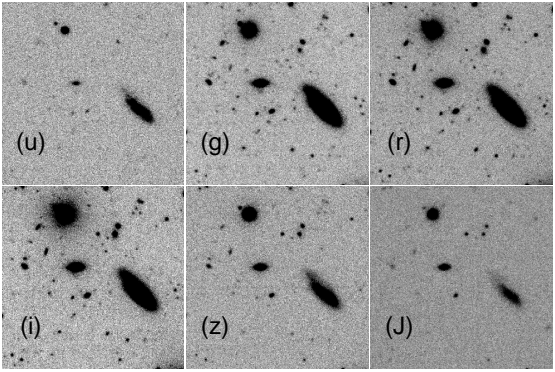}
\caption{Direct comparison between the SDSS optical data and the NEWFIRM
$J$-band data. The image size is $2\times2$ arcmin located at
00h50m01s+00d35m24s. The $J$-band image was chosen to have
a depth of of 20.2 mag (the average depth of our $J$-band images)
with a relatively good PSF of $1\farcs5$ (see Figure 15).
The SDSS co-adds are deeper than the NEWFIRM images.}
\end{figure}

Figure 17 shows the $z-J$ versus $i-z$ color-color diagram for point sources
brighter than $J=20$ mag in the region between $\rm R.A.=40\degr$ and 
$55\degr$. The point sources are selected using the distribution of object 
FWHMs in the $J$ band. Like the stellar locus in the $r-i$ versus $g-r$ 
diagram, the stellar locus in the $z-J$ versus $i-z$ diagram is very tight
in the brightest magnitude bin of $J<17$ mag. It becomes broader in fainter 
bins, as photometric errors increase. Since the $i$ and $z$-band data are much 
deeper than the $J$-band data, the photometric errors in the diagram are 
dominated by the $J$-band errors. In the faintest bin ($J>19$ mag), the 
stellar locus is not obvious any more due to large photometric errors and 
leakage of a large number of galaxies, as we saw in the faintest bin
in the $r-i$ versus $g-r$ diagram.
There is also a distinct clump of data points at $i-z=0.35$ and $z-J=1.45$ 
that are away from the stellar locus. It is not seen in the color-color
diagrams of stars in previous studies \citep[e.g.][]{fin00}. 
These sources are compact galaxies. The majority of them are classified as
extended sources by the SDSS. However, they were selected as stars by the
$J$-band data due to the poor seeing.

\section{SUMMARY AND DATA RELEASE}

In this paper we have introduced a new version of co-added images for the SDSS
Stripe 82. Stripe 82 covers 300 deg$^2$, and was repeatedly scanned 70--90 
times over roughly 10 years. These Stripe 82 images, when co-added, reach a 
much greater depth than do SDSS single-run data. We have described the details 
of the construction of our co-added images. In order to optimize the depth of 
the co-adds, we considered all available data in the SDSS archive and included 
as many images as possible, so that the marginal gain by adding more images is 
negligible. Each input image was properly processed and weighted based on 
PSF FWHM, sky transparency, and background noise. In particular, we performed
sky subtraction using a simple but efficient method that can properly deal
with the presence of large objects and strong background variation along the
drift scan direction. Our final products consist of 24060 science images and
their associated weight images and object catalogs. The weight images record 
relative weight at each pixel position. The catalogs were made with 
{\tt SExtractor}. Our co-adds reach more than two mag deeper than the deepest 
SDSS single-epoch images, and 0.3--0.5 mag deeper than the \citet{ann11} 
co-adds. They have good image quality with an average PSF FWHM of 
$\sim$$1\arcsec$ in the $r$, $i$, and $z$ bands. 
We have also presented $J$-band images obtained from NOAO NEWFIRM. These 
images cover roughly 90 deg$^2$ of Stripe 82 and have a depth of 20.0--20.5 
Vega magnitudes.

\begin{figure} %f17
%\epsscale{0.5}
\plotone{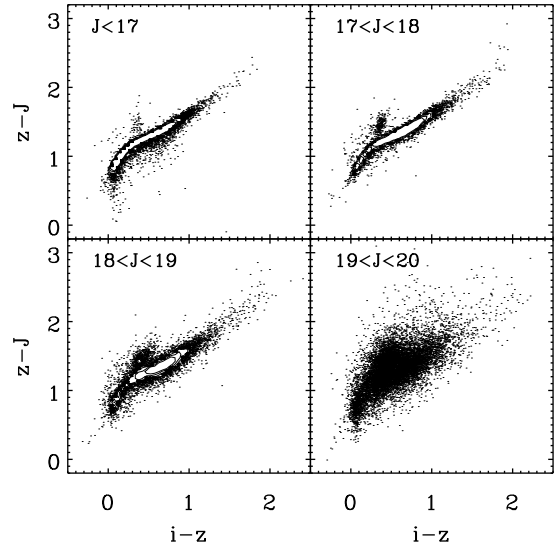}
\caption{The $z-J$ versus $i-z$ color-color diagram for point objects brighter
than $J=20$ mag in the region between $\rm R.A.=40\degr$ and $55\degr$. There
is a distinct clump of data points at $i-z=0.35$ and $z-J=1.45$ that are away
from the stellar locus. These sources are compact galaxies that were
identified as stars due to the poor image quality in the $J$-band data.}
\end{figure}

Our co-added images have many potential uses for studies of galaxies, quasars, 
and Galactic structure. The advantage of the data is the uniform coverage over 
300 deg$^2$ of the sky to a great depth. \citet{ann11} has listed many science 
opportunities, from Galactic dwarf stars to galaxy clusters. 
Here we briefly present two science cases not considered by 
\citet{ann11}. The first science case is high-redshift 
quasars, including quasars at $z\sim6$ and $z\sim5$. 
So far we have found 13 $z\sim6$ quasars down to $z\simeq22$ mag 
(10$\sigma$ detection) in Stripe 82 \citep{jiang08,jiang09}. Six of them are 
fainter than the depth of the \citet{ann11} co-adds. The critical part of the 
selection of these quasars is the $z-J$ versus $i-z$ color-color diagram, 
so $J$-band data such as our NEWFIRM data are also important
to find $z\sim6$ quasars. Our co-adds have also been used to find $z\sim5$
quasars \citep{mcg13}. These quasars are two magnitudes fainter than those
found in the SDSS single-epoch data, and allow us to probe the faint end of
the quasar luminosity function at this redshift.
The second science case is to find high-redshift Lyman-break galaxies (LBGs) 
at $z\ge3$ using the dropout technique. These distant galaxies are usually 
very faint and only found in deep small-area (several square degrees or 
smaller) surveys. Our co-adds are deep enough to find bright LBGs at $z\ge3$, 
given that the 5$\sigma$ depth in the $r$ band (24.6 mag) reaches the 
characteristic luminosity $L^{\ast}$ of $z=3$ LBGs. In particular, their sky 
coverage is large enough to find ultra-luminous LBGs (more than two 
magnitudes brighter than $L^{\ast}$) like the one reported by \citet{bian12}. 
Furthermore, the co-adds are likely more efficient to select bright LBGs
at $z\sim4$ based on the $g-r$ color, because the $g$ band is the deepest 
SDSS band. 

All our data products, including co-added science images and their 
associated weight images and object catalogs, are released on this web site, \\
http://das.sdss.org/ge/sample/stripe82/. 
The `Readme' file on the web site describes the structure of the dataset and 
how to download the data. The SDSS co-added data are under the folder or link
`sdss', and the NEWFIRM data are under the folder or link `newfirm'.
In addition, we will work on high-level catalogs with functions such as 
searchable tables and the cross-match of multiple bands. We will release
these products on the same web site when they are ready.

\acknowledgments

Support for this work was provided by NASA through Hubble Fellowship grant
HST-HF-51291.01 awarded by STScI, which is operated by the Association of 
Universities for Research in Astronomy, Inc., for NASA, under contract NAS 
5-26555. LJ also acknowledge support from the National Natural Science 
Foundation of China (NSFC) under grants 11003021 and 11373003.
XF and IDM acknowledge support from Packard Fellowship for Science and
Engineering and NSF grant AST 08-06861 and AST 11-07682. ZB acknowledges 
financial support from Princeton University to travel to Kitt Peak.

Funding for the SDSS and SDSS-II has been provided by the Alfred P. Sloan 
Foundation, the Participating Institutions, the National Science Foundation, 
the U.S. Department of Energy, the National Aeronautics and Space 
Administration, the Japanese Monbukagakusho, the Max Planck Society, and the 
Higher Education Funding Council for England. The SDSS Web Site is 
http://www.sdss.org/.

The SDSS is managed by the Astrophysical Research Consortium for the 
Participating Institutions. The participating institutions are the American 
Museum of Natural History, Astrophysical Institute Potsdam, University of 
Basel, University of Cambridge, Case Western Reserve University, University
of Chicago, Drexel University, Fermilab, the Institute for Advanced Study, the 
Japan Participation Group, Johns Hopkins University, the Joint Institute for 
Nuclear Astrophysics, the Kavli Institute for Particle Astrophysics and 
Cosmology, the Korean Scientist Group, the Chinese Academy of Sciences 
(LAMOST), Los Alamos National Laboratory, the Max-Planck-Institute for 
Astronomy (MPIA), the Max-Planck-Institute for Astrophysics (MPA), New Mexico 
State University, Ohio State University, University of Pittsburgh, University 
of Portsmouth, Princeton University, the United States Naval Observatory, and
the University of Washington.

{\it Facilities:}
\facility{Sloan},
\facility{Mayall (NEWFIRM)}

%\clearpage
\begin{deluxetable*}{ll}
\tablecaption{{\tt SExtractor} Configuration File}
\tablewidth{0pt}
\tablehead{\colhead{Keyword} & \colhead{Value}}
\startdata
CATALOG\_TYPE     & ASCII\_HEAD \\
DETECT\_MINAREA   & 4 \\
DETECT\_THRESH    & 2 \\
DEBLEND\_NTHRESH  & 16 \\
DEBLEND\_MINCONT  & 0.002 \\
WEIGHT\_TYPE      & MAP\_WEIGHT \\
PHOT\_APERTURES   & 6, 8, 10, 12, 20 \\
PHOT\_AUTOPARAMS  & 2.5, 3.5 \\
PHOT\_PETROPARAMS & 2.0, 3.5 \\
\enddata
\end{deluxetable*}

\begin{deluxetable*}{lll}
\tablecaption{{\tt SExtractor} Parameter File}
\tablewidth{0pt}
\tablehead{\colhead{Parameters} & \colhead{Units} & \colhead{Description}}
\startdata
NUMBER        &  ---  & Running object number \\
X\_IMAGE      & pixel & Object position along x \\
Y\_IMAGE      & pixel & Object position along y \\
ALPHA\_J2000  & deg   & R.A. (J2000) \\
DELTA\_J2000  & deg   & Decl. (J2000) \\
MAG\_APER     & mag   & Aperture mag \\
MAGERR\_APER  & mag   & Error for MAG\_APER \\
MAG\_AUTO     & mag   & Auto mag (Kron-like elliptical aperture mag) \\
MAGERR\_AUTO  & mag   & Error for MAG\_AUTO \\
MAG\_PETRO    & mag   & Petrosian mag (Petrosian-like elliptical aperture mag) \\
MAGERR\_PETRO & mag   & Error for MAG\_PETRO \\
A\_IMAGE      & pixel & Semi-major axis \\
B\_IMAGE      & pixel & Semi-minor axis \\
THETA\_IMAGE  & deg   & Position angle \\
ELLIPTICITY   & pixel & Ellipticity: 1-B\_IMAGE/A\_IMAGE \\
KRON\_RADIUS  &  ---  & Kron radius in units of A\_IMAGE \\
PETRO\_RADIUS &  ---  & Petrosian radius in units of A\_IMAGE \\
FWHM\_IMAGE   & pixel & FWHM assuming a Gaussian core \\
FLAGS         &  ---  & Internal extraction flags; sum of powers of 2 \\
\enddata
\end{deluxetable*}

\clearpage
\appendix
\section{SDSS Runs Used for Our Co-adds}
Table 1 shows the SDSS runs selected for our co-adds of Stripe 82. The first 
two columns are SDSS run numbers and MJD. Columns 3 and 4 show the starting 
and ending fields. Columns 5 and 6 show the starting and ending R.A. in 
degrees. The last column indicates that a run consists of south (S) or 
north (N) strips for Stripe 82.

\input{tab_app.tex}

\end{document}

%% file: tab_app.tex
%\clearpage
\LongTables
\begin{deluxetable}{rrrrrrr}
\tablecaption{SDSS Runs Used for Our Co-adds}
%\tablewidth{0pt}
\tablehead{\colhead{Run} & \colhead{MJD} & \colhead{$\rm Field_{start}$} &
   \colhead{$\rm Field_{end}$} & \colhead{$\rm R.A._{start}$} &
   \colhead{$\rm R.A._{end}$} & \colhead{Strip}}
\startdata
  125 &   51081 &   11 &  586 &  --9.45 &   76.57 &  S \\
 1033 &   51464 &   11 &  244 & --41.30 &  --6.42 &  N \\
 1056 &   51467 &   12 &  232 & --34.00 &  --1.05 &  S \\
 1752 &   51818 &   40 &  372 &   26.84 &   76.55 &  N \\
 1755 &   51819 &   74 &  683 & --45.23 &   45.95 &  S \\
 1894 &   51875 &   11 &  175 &   32.72 &   57.28 &  S \\
 2385 &   52075 &   11 &  100 & --52.53 & --39.21 &  N \\
 2570 &   52170 &  100 &  220 &   23.15 &   41.11 &  N \\
 2578 &   52171 &   60 &  210 &   37.45 &   59.90 &  N \\
 2579 &   52171 &   51 &  149 &   43.50 &   58.18 &  S \\
 2583 &   52172 &   30 &  254 & --52.53 & --18.99 &  S \\
 2585 &   52172 &   11 &   96 & --31.80 & --19.08 &  S \\
 2589 &   52173 &   81 &  305 &   27.35 &   60.88 &  N \\
 2649 &   52196 &   26 &  180 & --14.28 &    8.77 &  N \\
 2650 &   52196 &   11 &  175 &    5.23 &   29.78 &  N \\
 2659 &   52197 &   48 &  153 & --51.67 & --35.95 &  N \\
 2662 &   52197 &   18 &  436 & --39.59 &   22.99 &  N \\
 2677 &   52207 &   38 &  186 &    9.34 &   31.50 &  N \\
 2700 &   52224 &   22 &  276 &   23.43 &   61.45 &  N \\
 2708 &   52225 &   25 &  270 & --12.48 &   24.19 &  N \\
 2709 &   52225 &   25 &  261 &   23.54 &   58.87 &  S \\
 2728 &   52231 &  150 &  628 & --39.35 &   32.21 &  N \\
 2738 &   52234 &   18 &  321 &   14.96 &   60.32 &  N \\
 2768 &   52253 &   26 &  237 & --15.92 &   15.67 &  N \\
 2820 &   52261 &   22 &  265 &   23.36 &   59.74 &  N \\
 2855 &   52282 &   11 &   66 &   20.81 &   29.04 &  N \\
 2861 &   52283 &   13 &  207 &   34.24 &   63.28 &  N \\
 2873 &   52287 &   55 &  310 &   21.85 &   60.02 &  N \\
 2886 &   52288 &   53 &  308 &   21.56 &   59.73 &  S \\
 3325 &   52522 &   11 &  506 & --14.48 &   59.63 &  S \\
 3355 &   52551 &   11 &  275 &   20.44 &   59.96 &  S \\
 3360 &   52552 &   11 &  522 & --52.40 &   24.10 &  S \\
 3362 &   52552 &   11 &  235 &   21.52 &   55.06 &  N \\
 3384 &   52557 &   18 &  779 & --52.55 &   61.38 &  N \\
 3388 &   52558 &   11 &  723 & --46.17 &   60.42 &  S \\
 3427 &   52576 &   28 &  145 & --47.80 & --30.29 &  S \\
 3430 &   52576 &   13 &  117 &   22.11 &   37.68 &  S \\
 3434 &   52577 &   28 &  575 & --48.61 &   33.27 &  S \\
 3437 &   52578 &   18 &  500 & --48.68 &   23.47 &  N \\
 3438 &   52578 &   11 &  200 &   31.68 &   59.98 &  S \\
 3460 &   52585 &   25 &  275 &   22.57 &   59.99 &  S \\
 3461 &   52585 &   11 &  118 &   43.83 &   59.85 &  N \\
 3465 &   52586 &   11 &  359 & --33.12 &   18.98 &  S \\
 4128 &   52908 &   11 &  524 & --16.77 &   60.04 &  N \\
 4136 &   52909 &   11 &  215 &   28.94 &   59.48 &  S \\
 4145 &   52910 &   11 &  514 & --14.82 &   60.48 &  S \\
 4153 &   52911 &   11 &  182 & --15.18 &   10.42 &  N \\
 4157 &   52912 &   11 &  276 &   20.29 &   59.96 &  N \\
 4184 &   52929 &   31 &  312 & --53.69 & --11.62 &  N \\
 4187 &   52930 &   11 &  113 & --50.67 & --35.40 &  S \\
 4188 &   52930 &   11 &  154 & --14.79 &    6.62 &  N \\
 4192 &   52931 &   11 &  502 & --51.61 &   21.89 &  S \\
 4198 &   52934 &   11 &  761 & --52.51 &   59.77 &  N \\
 4203 &   52935 &   33 &  806 & --55.66 &   60.06 &  S \\
 4207 &   52936 &   11 &  772 & --53.95 &   59.97 &  N \\
 4247 &   52959 &   11 &  218 & --14.60 &   16.38 &  S \\
 4253 &   52962 &   11 &  182 & --14.53 &   11.07 &  N \\
 4263 &   52963 &   11 &  467 & --15.64 &   52.62 &  S \\
 4288 &   52971 &   11 &  178 &   20.32 &   45.32 &  S \\
 4797 &   53243 &   11 &  190 & --52.48 & --25.68 &  N \\
 4849 &   53270 &   11 &  941 & --66.39 &   72.84 &  N \\
 4858 &   53272 &   18 &  749 & --50.34 &   59.10 &  N \\
 4868 &   53286 &   11 &  619 & --29.48 &   61.54 &  N \\
 4874 &   53288 &   11 & 1000 & --61.35 &   86.71 &  N \\
 4894 &   53294 &   11 &  207 & --32.41 &  --3.07 &  N \\
 4895 &   53294 &   11 &  497 &  --3.74 &   69.02 &  N \\
 4899 &   53296 &   11 &  360 &    9.26 &   61.50 &  N \\
 4905 &   53298 &   11 &  471 &    1.43 &   70.30 &  N \\
 4917 &   53302 &   11 &  786 & --64.54 &   51.48 &  N \\
 4927 &   53312 &   11 &  760 & --51.03 &   61.10 &  N \\
 4930 &   53313 &   11 &  395 & --56.96 &    0.53 &  S \\
 4933 &   53314 &   11 &  775 & --52.65 &   61.72 &  N \\
 4948 &   53319 &   66 &  350 &   18.44 &   60.96 &  N \\
 5042 &   53351 &   11 &  281 &   19.47 &   59.89 &  S \\
 5052 &   53352 &   11 &  272 & --14.61 &   24.46 &  S \\
 5566 &   53616 &   11 &  622 & --32.57 &   58.91 &  N \\
 5582 &   53622 &   11 &  759 & --54.58 &   57.41 &  S \\
 5590 &   53623 &   11 &  480 & --59.65 &   10.56 &  N \\
 5597 &   53625 &   11 &  313 & --63.64 & --18.43 &  S \\
 5603 &   53626 &  110 &  850 & --50.62 &   60.17 &  N \\
 5607 &   53627 &   26 &   92 & --60.62 & --50.74 &  S \\
 5607 &   53627 &  700 &  831 &   40.28 &   59.90 &  S \\
 5610 &   53628 &   11 &  877 & --65.69 &   63.96 &  N \\
 5619 &   53634 &   96 &  840 & --50.68 &   60.71 &  S \\
 5622 &   53635 &   11 &  848 & --63.45 &   61.86 &  N \\
 5628 &   53636 &   11 &  449 & --63.58 &    1.99 &  S \\
 5633 &   53637 &   11 &  659 & --60.56 &   36.45 &  N \\
 5637 &   53638 &   11 &  562 & --20.57 &   61.92 &  S \\
 5642 &   53639 &   11 &  486 &  --9.67 &   61.44 &  N \\
 5646 &   53640 &  350 &  905 & --13.85 &   69.24 &  S \\
 5654 &   53641 &   11 &  124 & --69.96 & --53.02 &  N \\
 5658 &   53641 &   11 &  267 &   16.23 &   54.56 &  N \\
 5665 &   53643 &   11 &  103 & --67.97 & --54.20 &  S \\
 5666 &   53643 &   11 &  148 &   41.40 &   61.91 &  S \\
 5670 &   53644 &   64 &  869 & --55.53 &   64.98 &  N \\
 5675 &   53645 &   11 &  130 & --59.56 & --41.75 &  S \\
 5681 &   53646 &   11 &  208 &   23.38 &   52.88 &  S \\
 5698 &   53648 &  230 &  357 & --28.83 &  --9.81 &  S \\
 5702 &   53649 &   90 &  152 & --55.70 & --46.42 &  N \\
 5709 &   53654 &  110 &  611 & --51.60 &   23.40 &  N \\
 5713 &   53655 &   11 &  738 & --67.36 &   41.47 &  S \\
 5719 &   53656 &   11 &  445 & --62.63 &    2.34 &  N \\
 5731 &   53657 &   11 &  276 &   21.32 &   60.99 &  N \\
 5732 &   53657 &   11 &  103 &   47.26 &   61.03 &  S \\
 5744 &   53663 &   11 &  595 & --29.60 &   57.83 &  N \\
 5745 &   53663 &   11 &  157 &   33.23 &   55.10 &  S \\
 5754 &   53664 &   11 &  776 & --56.65 &   57.88 &  S \\
 5760 &   53665 &   11 &  214 &   25.28 &   55.68 &  S \\
 5763 &   53666 &   29 &  426 & --55.49 &    3.94 &  S \\
 5765 &   53666 &   11 &  361 &    2.45 &   54.86 &  N \\
 5770 &   53668 &   11 &  769 & --55.54 &   57.94 &  N \\
 5771 &   53668 &   11 &  196 &   33.24 &   60.94 &  S \\
 5776 &   53669 &   11 &  792 & --58.95 &   57.97 &  S \\
 5777 &   53669 &   11 &  234 &   24.56 &   57.96 &  N \\
 5781 &   53670 &   11 &  766 & --55.16 &   57.86 &  N \\
 5782 &   53670 &   11 &  199 &   32.80 &   60.96 &  S \\
 5786 &   53671 &   11 &  662 & --35.63 &   61.85 &  S \\
 5792 &   53673 &   11 &  808 & --61.40 &   57.91 &  N \\
 5797 &   53674 &   11 &  785 & --57.97 &   57.90 &  S \\
 5800 &   53675 &   37 &  793 & --54.56 &   58.62 &  N \\
 5807 &   53676 &   11 &  715 & --47.52 &   57.89 &  S \\
 5808 &   53676 &   11 &   68 &   49.42 &   57.96 &  N \\
 5813 &   53677 &   11 &  737 & --64.09 &   44.59 &  N \\
 5820 &   53679 &   24 &  708 & --42.44 &   59.96 &  S \\
 5823 &   53680 &   11 &  813 & --59.06 &   61.01 &  N \\
 5836 &   53681 &   11 &  816 & --59.53 &   61.00 &  S \\
 5842 &   53683 &  600 &  807 &   29.91 &   60.91 &  N \\
 5847 &   53684 &  600 &  846 &   25.71 &   62.54 &  S \\
 5865 &   53686 &   11 &  176 & --42.82 & --18.11 &  N \\
 5866 &   53686 &   11 &  295 &   18.36 &   60.88 &  N \\
 5870 &   53687 &   11 &  386 & --62.16 &  --6.01 &  S \\
 5871 &   53687 &   11 &  210 &   31.54 &   61.33 &  S \\
 5878 &   53693 &   22 &  828 & --59.98 &   60.68 &  N \\
 5882 &   53694 &   28 &  835 & --60.11 &   60.70 &  S \\
 5889 &   53696 &   62 &  180 &   44.01 &   61.68 &  S \\
 5895 &   53697 &   25 &  832 & --60.11 &   60.71 &  S \\
 5898 &   53698 &  240 &  389 & --30.49 &  --8.18 &  N \\
 5898 &   53698 &   11 &  219 & --64.78 & --33.63 &  N \\
 5898 &   53698 &  430 &  714 &  --2.04 &   40.48 &  N \\
 5902 &   53699 &   11 &  549 & --61.77 &   18.78 &  N \\
 5902 &   53699 &  650 &  829 &   33.90 &   60.71 &  N \\
 5905 &   53700 &   58 &  865 & --59.99 &   60.84 &  S \\
 5909 &   53702 &   50 &  389 & --57.55 &  --6.80 &  N \\
 5915 &   53703 &   16 &  236 & --60.01 & --27.07 &  N \\
 5918 &   53704 &   57 &  827 & --54.46 &   60.82 &  N \\
 5924 &   53705 &   25 &  833 & --60.06 &   60.91 &  S \\
 6281 &   53974 &   11 &  159 & --56.67 & --34.52 &  N \\
 6283 &   53974 &   11 &  172 & --11.54 &   12.57 &  N \\
 6287 &   53975 &   11 &  810 & --58.63 &   61.00 &  S \\
 6293 &   53977 &   11 &   51 &    7.46 &   13.45 &  N \\
 6313 &   53989 &   11 &  124 & --62.75 & --45.84 &  N \\
 6314 &   53989 &   11 &  735 & --47.24 &   61.15 &  N \\
 6330 &   53990 &   11 &  189 &    1.32 &   27.97 &  S \\
 6348 &   53993 &   11 &  238 & --31.04 &    2.94 &  S \\
 6349 &   53993 &   11 &   96 &   26.26 &   38.98 &  S \\
 6353 &   53994 &   11 &   97 & --61.64 & --48.77 &  S \\
 6355 &   53994 &   11 &  449 &  --5.63 &   59.95 &  S \\
 6360 &   53995 &   23 &  325 & --59.98 & --14.77 &  N \\
 6362 &   53995 &   11 &  122 &    9.28 &   25.90 &  N \\
 6363 &   53995 &   11 &  106 &   25.82 &   40.04 &  N \\
 6367 &   53996 &   11 &  737 & --47.67 &   61.02 &  S \\
 6370 &   53997 &   11 &  271 & --64.78 & --25.85 &  N \\
 6373 &   53997 &   11 &  181 & --17.07 &    8.39 &  N \\
 6374 &   53997 &   11 &   97 &   38.31 &   51.18 &  N \\
 6377 &   53998 &   11 &  560 & --66.44 &   15.75 &  S \\
 6383 &   54000 &   11 &  587 & --32.73 &   53.50 &  N \\
 6391 &   54003 &   11 &   66 & --62.60 & --54.37 &  N \\
 6400 &   54005 &   11 &  209 & --60.62 & --30.98 &  N \\
 6401 &   54005 &   11 &   72 & --50.75 & --41.62 &  S \\
 6402 &   54005 &   11 &  256 & --24.77 &   11.91 &  S \\
 6404 &   54005 &   11 &  250 &   25.12 &   60.90 &  S \\
 6408 &   54006 &   55 &  130 & --59.99 & --48.77 &  S \\
 6409 &   54006 &  240 &  409 &    1.65 &   26.95 &  N \\
 6412 &   54007 &   11 &  242 & --58.76 & --24.18 &  N \\
 6414 &   54007 &   11 &  550 & --25.78 &   54.91 &  N \\
 6417 &   54008 &   11 &  753 & --60.73 &   50.35 &  S \\
 6418 &   54008 &   11 &   61 &   53.36 &   60.85 &  N \\
 6421 &   54009 &   11 &  834 & --62.37 &   60.84 &  N \\
 6422 &   54009 &   11 &  137 &   37.24 &   56.10 &  S \\
 6425 &   54010 &   11 &  828 & --61.39 &   60.92 &  S \\
 6430 &   54011 &   11 &  836 & --62.65 &   60.86 &  N \\
 6433 &   54012 &   11 &  852 & --65.02 &   60.88 &  S \\
 6435 &   54012 &   11 &  133 &   36.24 &   54.50 &  N \\
 6441 &   54019 &   11 &  467 & --61.63 &    6.64 &  N \\
 6444 &   54019 &  158 &  171 &   43.29 &   45.24 &  N \\
 6447 &   54020 &   11 &  724 & --66.64 &   40.10 &  S \\
 6448 &   54020 &   11 &  196 &   39.23 &   66.93 &  S \\
 6450 &   54021 &   11 &  465 &  --6.59 &   61.38 &  N \\
 6453 &   54022 &   11 &  399 & --62.40 &  --4.32 &  S \\
 6458 &   54024 &   11 &  428 &  --6.58 &   55.85 &  S \\
 6461 &   54025 &   11 &  824 & --60.71 &   61.01 &  N \\
 6464 &   54026 &   11 &  171 & --68.67 & --44.72 &  S \\
 6468 &   54028 &   11 &  449 & --62.60 &    2.96 &  S \\
 6471 &   54028 &   11 &  360 &    9.68 &   61.93 &  S \\
 6474 &   54029 &   11 &  639 & --62.60 &   31.41 &  N \\
 6476 &   54029 &   11 &  214 &   30.71 &   61.11 &  N \\
 6479 &   54030 &   11 &  665 & --63.08 &   34.82 &  S \\
 6480 &   54030 &   11 &  185 &   33.34 &   59.39 &  S \\
 6484 &   54031 &   11 &  837 & --62.70 &   60.96 &  N \\
 6488 &   54032 &   11 &  277 & --66.65 & --26.83 &  S \\
 6494 &   54034 &   17 &  209 & --61.51 & --32.77 &  S \\
 6501 &   54035 &   11 &  639 & --32.41 &   61.62 &  S \\
 6504 &   54036 &   11 &  845 & --62.99 &   61.86 &  N \\
 6508 &   54037 &   11 &  831 & --62.62 &   60.14 &  S \\
 6513 &   54039 &   11 &  837 & --62.63 &   61.03 &  N \\
 6518 &   54040 &   11 &  837 & --62.67 &   60.99 &  S \\
 6524 &   54041 &   11 &  382 &    6.25 &   61.80 &  N \\
 6525 &   54041 &   11 &  158 &   39.27 &   61.28 &  S \\
 6530 &   54047 &   11 &  475 &  --7.83 &   61.64 &  S \\
 6533 &   54048 &   11 &  714 & --65.34 &   39.91 &  N \\
 6534 &   54048 &   11 &  171 &   38.85 &   62.80 &  N \\
 6537 &   54049 &   11 &  533 & --17.23 &   60.93 &  S \\
 6542 &   54050 &   11 &  433 & --62.68 &    0.49 &  S \\
 6545 &   54050 &   11 &  142 &   41.37 &   60.98 &  S \\
 6548 &   54051 &   11 &  289 & --63.62 & --22.01 &  N \\
 6552 &   54052 &   11 &  843 & --62.59 &   61.97 &  N \\
 6555 &   54053 &   11 &  661 & --66.16 &   31.15 &  S \\
 6556 &   54053 &   11 &  229 &   28.52 &   61.16 &  S \\
 6559 &   54054 &   11 &  523 & --62.60 &   14.05 &  N \\
 6564 &   54055 &   11 &  636 & --63.24 &   30.33 &  N \\
 6565 &   54055 &   11 &  230 &   28.32 &   61.11 &  N \\
 6568 &   54056 &   11 &  850 & --64.69 &   60.92 &  S \\
 6571 &   54057 &   11 &  240 & --53.61 & --19.32 &  S \\
 6577 &   54058 &   11 &  846 & --63.58 &   61.43 &  N \\
 6580 &   54059 &   11 &  859 & --65.45 &   61.51 &  S \\
 6584 &   54060 &   11 &  703 & --42.68 &   60.93 &  N \\
 6590 &   54061 &   11 &  336 & --62.18 & --13.52 &  S \\
 6592 &   54061 &   11 &  336 &   12.22 &   60.88 &  S \\
 6596 &   54062 &   11 &  837 & --62.77 &   60.90 &  S \\
 6600 &   54063 &   11 &  676 & --63.69 &   35.87 &  N \\
 6604 &   54064 &   11 &  239 &   27.39 &   61.53 &  S \\
 6609 &   54065 &   11 &  195 &   33.33 &   60.88 &  N \\
 6615 &   54068 &   11 &  324 & --63.64 & --16.78 &  S \\
 6618 &   54068 &   11 &  430 &  --1.66 &   61.07 &  S \\
 6920 &   54346 &   74 &  621 & --60.22 &   21.64 &  N \\
 6921 &   54346 &   11 &  216 &   24.20 &   54.89 &  N \\
 6930 &   54347 &   11 &  506 & --13.59 &   60.52 &  S \\
 6933 &   54348 &   35 &  342 & --60.04 & --14.08 &  S \\
 6934 &   54348 &   11 &  482 & --14.56 &   55.96 &  N \\
 6947 &   54355 &   11 &   50 &   32.40 &   38.24 &  N \\
 6951 &   54356 &   11 &  224 & --52.55 & --20.67 &  N \\
 6955 &   54357 &   11 &  861 & --66.25 &   61.01 &  S \\
 6958 &   54358 &   65 &  873 & --60.07 &   60.90 &  N \\
 6961 &   54359 &   55 &   87 & --39.54 & --34.75 &  N \\
 6962 &   54359 &   24 &  273 & --34.73 &    2.55 &  N \\
 6963 &   54359 &   11 &  189 &  --0.67 &   25.98 &  N \\
 6964 &   54359 &   11 &  123 &   14.59 &   31.35 &  S \\
 6976 &   54362 &  170 &  709 & --19.72 &   60.97 &  S \\
 6981 &   54365 &   41 &  315 & --60.05 & --19.04 &  S \\
 6982 &   54365 &   11 &  603 & --27.71 &   60.92 &  N \\
 6985 &   54366 &   11 &  200 & --63.51 & --35.22 &  N \\
 7003 &   54373 &   11 &  133 & --50.45 & --32.19 &  N \\
 7006 &   54373 &   27 &  267 & --34.23 &    1.70 &  N \\
 7013 &   54376 &   11 &  128 & --25.63 &  --8.11 &  S \\
 7016 &   54376 &   11 &  335 &   12.31 &   60.82 &  S \\
 7018 &   54377 &   34 &   96 & --59.74 & --50.46 &  S \\
 7024 &   54379 &   35 &  542 & --60.03 &   15.87 &  S \\
 7033 &   54381 &   11 &  156 & --51.65 & --29.94 &  N \\
 7034 &   54381 &   11 &  536 & --17.61 &   60.99 &  N \\
 7037 &   54382 &   11 &  362 & --63.66 & --11.11 &  N \\
 7038 &   54382 &   11 &  449 &  --4.73 &   60.85 &  S \\
 7043 &   54383 &   11 &  168 & --57.00 & --33.49 &  N \\
 7047 &   54384 &   11 &  271 &    5.37 &   44.29 &  N \\
 7051 &   54385 &   11 &  835 & --62.45 &   60.91 &  S \\
 7054 &   54386 &   11 &  812 & --58.56 &   61.36 &  N \\
 7057 &   54387 &   11 &  832 & --62.04 &   60.88 &  S \\
 7060 &   54388 &   26 &  832 & --59.52 &   61.15 &  N \\
 7069 &   54390 &   11 &   82 & --53.65 & --43.02 &  S \\
 7071 &   54390 &   11 &  338 &   12.33 &   61.29 &  S \\
 7074 &   54391 &   11 &  343 &   11.22 &   60.93 &  S \\
 7076 &   54392 &   11 &  529 & --62.66 &   14.89 &  S \\
 7077 &   54392 &   11 &  302 &   17.28 &   60.85 &  N \\
 7080 &   54393 &   22 &  526 & --60.12 &   15.33 &  N \\
 7081 &   54393 &   36 &  321 &   18.27 &   60.94 &  N \\
 7084 &   54394 &   11 &  836 & --62.68 &   60.83 &  S \\
 7092 &   54396 &   11 &  256 & --65.62 & --28.94 &  N \\
 7095 &   54396 &   11 &  368 &    7.70 &   61.15 &  N \\
 7096 &   54396 &   11 &  322 &   14.67 &   61.24 &  S \\
 7101 &   54402 &   11 &  803 & --57.65 &   60.94 &  S \\
 7106 &   54403 &   11 &  851 & --64.66 &   61.09 &  N \\
 7110 &   54404 &   11 &  284 & --62.95 & --22.08 &  S \\
 7111 &   54404 &   11 &   65 & --21.72 & --13.64 &  N \\
 7112 &   54404 &   11 &  476 &  --8.59 &   61.03 &  N \\
 7117 &   54405 &   11 &  382 & --16.66 &   38.88 &  N \\
 7121 &   54406 &   11 &  836 & --62.60 &   60.91 &  N \\
 7127 &   54408 &   11 &  870 & --67.54 &   61.05 &  S \\
 7130 &   54409 &  702 &  840 &   40.24 &   60.90 &  N \\
 7133 &   54409 &   11 &  310 & --63.74 & --18.98 &  S \\
 7136 &   54411 &   11 &  741 & --48.27 &   61.02 &  S \\
 7140 &   54412 &   11 &  556 & --62.85 &   18.73 &  N \\
 7142 &   54412 &   11 &  309 &   16.37 &   60.99 &  N \\
 7145 &   54413 &   11 &  846 & --63.92 &   61.09 &  S \\
 7150 &   54415 &   11 &  189 & --62.61 & --35.96 &  N \\
 7151 &   54415 &   11 &   71 & --17.24 &  --8.26 &  S \\
 7152 &   54415 &  215 &  474 &   21.35 &   60.12 &  S \\
 7155 &   54416 &   11 &  856 & --65.61 &   60.89 &  N \\
 7158 &   54417 &   11 &  443 &  --3.63 &   61.05 &  S \\
 7161 &   54418 &   11 &  855 & --65.42 &   60.94 &  S \\
 7167 &   54420 &   11 &  399 & --52.24 &    5.85 &  N \\
 7170 &   54421 &   11 &  838 & --62.45 &   61.37 &  N \\
 7173 &   54422 &   11 &  836 & --62.62 &   60.89 &  S \\
 7176 &   54423 &   12 &  261 & --62.39 & --25.12 &  N \\
 7177 &   54423 &   11 &  487 & --10.37 &   60.89 &  N \\
 7182 &   54424 &   11 &  349 & --64.57 & --13.97 &  S \\
 7183 &   54424 &   11 &  362 &    8.33 &   60.88 &  S \\
 7188 &   54425 &   14 &  410 & --66.16 &  --6.87 &  N \\
 7202 &   54433 &   11 &  709 & --43.63 &   60.89 &  N \\
\enddata
%\tablecomments{$\rm R.A._{start}$ and $\rm R.A._{end}$ are in units of degree.}
\end{deluxetable}